\documentclass[runningheads]{llncs}
\usepackage{graphicx}
\usepackage{amsmath,amssymb}
\usepackage{cite}
\usepackage[french,english]{babel}
\usepackage{hyperref}
\usepackage{microtype}
\usepackage{caption}
\usepackage{subcaption}
\usepackage{tabularx}
\usepackage{multirow}
\usepackage{colortbl}
\usepackage{xcolor}
\usepackage{numprint}

\newcommand{\lmorph}{\mathcal{L}\text{Morph}}
\newcommand{\smorph}{\mathcal{S}\text{Morph}}

\begin{document}
\title{Going beyond p-convolutions to learn grayscale morphological operators}
\titlerunning{Going beyond p-convolutions}
%
\author{Alexandre Kirszenberg\inst{1} \and
Guillaume Tochon\inst{1} \and \'{E}lodie Puybareau\inst{1} \and
Jesus Angulo\inst{2}}
\authorrunning{A. Kirszenberg et al.}
%
\institute{EPITA Research and Development Laboratory (LRDE),
Le Kremlin-Bic\^{e}tre, France\\
\email{surname.name@lrde.epita.fr}\\
\and
Centre for Mathematical Morphology, Mines ParisTech, PSL Research University, France\\
\email{jesus.angulo@mines-paristech.fr}
}
\maketitle              
\begin{abstract}
Integrating mathematical morphology operations within deep neural networks has been subject to increasing attention lately. However, replacing standard convolution layers with erosions or dilations is particularly challenging because the $\min$ and $\max$ operations are not differentiable. 
Relying on the asymptotic behavior of the counter-harmonic mean, p-convolutional layers were proposed as a possible workaround to this issue since they can perform pseudo-dilation or pseudo-erosion operations (depending on the value of their inner parameter $p$), and very promising results were reported.
In this work, we present two new morphological layers based on the same principle as the p-convolutional layer while circumventing its principal drawbacks, and demonstrate their potential interest in further implementations within deep convolutional neural network architectures.

\end{abstract}
\section{Introduction}\label{sec:intro}

Mathematical morphology deals with the non-linear filtering of images~\cite{serra1983image}.
The elementary operations of mathematical morphology amount to computing the minimum (for the erosion) or maximum (for the dilation) of all pixel values within a neighborhood of some given shape and size (the structuring element) of the pixel under study. Combining those elementary operations, one can define more advanced (but still non-linear) filters, such as openings and closings, which have many times proven to be successful at various image processing tasks such as filtering, segmentation or edge detection~\cite{soille2013morphological}. However, deriving the optimal combination of operations and the design (shape and size) of their respective structuring element is generally done in a tedious and time-consuming trial-and-error fashion. Thus, delegating the automatic identification of the right sequence of operations to use and their structuring element to some machine learning technique is an appealing strategy.\\
On the other hand, artificial neural networks are composed of units (or neurons) connected to each other and organized in layers. The output of each neuron is expressed as the linear combination of its inputs weighed by trainable weights, potentially mapped by a non-linear activation function~\cite{hassoun1995fundamentals}. Convolutional neural networks (CNNs) work in a similar fashion, replacing neurons with convolutional filters~\cite{lecun2015deep}.\\
Because of the similarity between their respective operations, there has been an increasing interest in past years to integrate morphological operations within the framework of neural networks, and two major lines of research have emerged.
The first one, tracing back to the end of the 80s, replaces the multiplication and addition of linear perceptron units with addition and maximum~\cite{wilson1989morphological,davidson1990theory,ritter1996introduction}, resulting in so-called non-linear morphological perceptrons~\cite{sussner1998morphological} (see~\cite{charisopoulos2017morphological,zhang2019max} for recent works in this domain).
The second line, mainly motivated by the rise of deep CNNs, explores the integration of elementary morphological operations in such networks to automatically learn their optimal shape and weights, the major issue being that the $\min$ and $\max$ operations are not differentiable.
A first workaround is to replace them by smooth differentiable approximations, making them suited to the conventional gradient descent learning approach via back-propagation~\cite{lecun2015deep}. In their seminal work, Masci \textit{et al.}~\cite{masci2013learning} relied on some properties of the counter-harmonic mean~\cite{bullen2013handbook} (CHM) to provide p-convolutional ($PConv$) layers, the value of the trainable parameter $p$ dictating which of the elementary morphological operation the layer ultimately mimicks.
The CHM was also used as an alternative to the standard max-pooling layer in classical CNN architectures~\cite{mellouli2017morph}. LogSumExp functions (also known as multivariate softplus) were proposed as replacements of min and max operations to learn binary~\cite{shih2019development} and grayscale~\cite{shen2019deep} structuring elements.
An alternative approach was followed in~\cite{mondal2020image,franchi2020deep}: the non-linear morphological operations remained unchanged, and the backpropagation step was instead adapted to handle them in the same way the classical max-pooling layer is handled in standard CNNs.
Finally, morphological operations were recreated and optimized as combinations of depthwise and pointwise convolution with depthwise pooling~\cite{nogueira2019introduction}.\\
Looking at all recently proposed approaches (apart from~\cite{masci2013learning}, all other aforementionned works date back to no later than 2017) and the diversity of their evaluation (image classification on MNIST database~\cite{lecun1998mnist}, image denoising and restoration, edge detection and so on), it seems to us that the magical formula for integrating morphological operations within CNNs has yet to be derived. For this reason, we would like to draw the attention in this work back to the $PConv$ layer proposed in~\cite{masci2013learning}. As a matter of fact, very promising results were reported but never investigated further. Relying on the CHM framework, we propose two possible extensions to the $PConv$ layer, for which we demonstrate potential interest in further implementations within deep neural network architectures.\\
In section~\ref{sec:p-conv}, we review the work on p-convolutions~\cite{masci2013learning}, presenting their main advantages, properties and limitations. In section~\ref{sec:contrib}, we propose two new morphological layers, namely the $\lmorph$ layer (also based on the CHM) and the $\smorph$ layer (based on the regularized softmax). Both proposed layers are compatible with grayscale mathematical morphology and nonflat structuring elements.
In Section~\ref{sec:xp}, we showcase some results from our implementations, and proceed to compare these results to those of the p-convolution layer.
Finally, Section~\ref{sec:concl} provides a conclusion and some perspectives from our contributions.
\vspace{-5pt}

\section{P-convolutions: definitions, properties and pitfalls}\label{sec:p-conv}

In this section, we detail the notion of p-convolution as presented in~\cite{masci2013learning}.
\vspace{-5pt}

\subsection{Grayscale mathematical morphology}\label{subsec:grayscale_morpho}

In mathematical morphology, an image is classically represented as a 2D function $f : E \rightarrow \mathbb{R}$ with $x \in E$ being the pixel coordinates in the 2D grid $E \subseteq \mathbb{Z}^2$ and $f(x) \in \mathbb{R}$ being the pixel value.
In grayscale mathematical morphology, \textit{i.e.} when both the image $f$ and the structuring element $b$ are real-valued (and not binary), erosion $f \ominus b$ and dilation $f \oplus b$ operations can be written as:
\begin{align}
(f \ominus b)(x) = \inf_{y \in E} \left\{f(y) - b(x - y)\right\}\label{eq:grayscale_erode}\\
(f \oplus b)(x) = \sup_{y \in E} \left\{f(y) + b(x - y)\right\}\label{eq:grayscale_dilate}
\end{align}
This formalism also encompasses the use of flat (binary) structuring elements, which are then written as\vspace{-8pt}
\begin{equation}\label{eq:flat_structuringelement}
b(x) =  \begin{cases} 0 &\text{ if } x \in B \\ -\infty & \text{ otherwise}\end{cases},
\end{equation}
where $B \subseteq E$ is the support of the structuring function $b$.
\vspace{-5pt}

\subsection{The counter-harmonic mean and the p-convolution}\label{subsec:CHM_pconv}

Let $p \in \mathbb{R}$.
The counter-harmonic mean (CHM) of order $p$ of a given non negative vector $\mathbf{x} = (x_1,\dots,x_n) \in (\mathbb{R}^+)^n$ with non negative weights $\mathbf{w} = (w_1,\dots,w_n) \in (\mathbb{R}^+)^n$ is defined as\vspace{-5pt}
\begin{equation}\label{eq:CHM}
\displaystyle CHM(\mathbf{x},\mathbf{w},p) = \frac{\sum_{i=1}^n w_i x_i^p}{\sum_{i=1}^{n} w_i x_i^{p-1}}\ .
\end{equation}
The CHM is also known as the Lehmer mean~\cite{bullen2013handbook}.
Asymptotically, one has $\lim_{p \rightarrow +\infty} CHM(\mathbf{x},\mathbf{w},p) = \sup_i x_i$ and $\lim_{p \rightarrow -\infty} CHM(\mathbf{x},\mathbf{w},p) = \inf_i x_i$.\\
The p-convolution of an image $f$ at pixel $x$ for a given (positive) convolution kernel $w : W \subseteq E \rightarrow \mathbb{R}^+$ is defined as:
\begin{equation}\label{eq:pconv}\small
PConv(f, w, p)(x) = (f *_p w)(x) = \frac{(f^{p+1} * w)(x)}{(f^p * w)(x)} = \frac{\sum_{y \in W(x)} f^{p+1}(y) w(x-y)}{\sum_{y \in W(x)} f^{p}(y) w(x-y)}
\end{equation}
\normalsize
where $f^p(x)$ denotes the pixel value $f(x)$ raised at the power of $p$, $W(x)$ is the spatial support of kernel $w$ centered at $x$, and the scalar $p$ controls the type of operation to perform.\\
Based on the asymptotic properties of the CHM, the morphological behavior of the $PConv$ operation with respect to $p$ has notably been studied in~\cite{angulo2010pseudo}.
More precisely, when $p > 0$ (resp. $p<0$), the operation is a pseudo-dilation (resp. pseudo-erosion), and when $p\to \infty$ (resp. $-\infty$), the largest (resp. smallest) pixel value in the local neighborhood $W(x)$ of pixel $x$ dominates the weighted sum~\eqref{eq:pconv} and the $PConv(f, w, p)(x)$ acts as a non-flat grayscale dilation (resp. a non-flat grayscale erosion) with the structuring function $b(x) = \frac1p log (w(x))$:
\begin{align}
\lim_{p \rightarrow +\infty} (f *_p w)(x) = \sup_{y \in W(x)} \left\{ f(y) + \frac1p \log \left(w(x - y)\right)\right\}\label{eq:pconv_dilate}\\
\lim_{p \rightarrow -\infty} (f *_p w)(x) = \inf_{y \in W(x)} \left\{ f(y) - \frac1p \log \left(w(x - y)\right)\right\}\label{eq:pconv_erode}
\end{align}
In practice, equations~\eqref{eq:pconv_dilate} and~\eqref{eq:pconv_erode} hold true for $\vert p \vert > 10$. The flat structuring function~\eqref{eq:flat_structuringelement} can be recovered by using constant weight kernels, \textit{i.e.}, $w(x) = 1$ if $x \in W$
and $w(x) = 0$ if $x \not\in W$ and $\vert p \vert \gg 0$.
As stated in~\cite{masci2013learning}, the $PConv$ operation is differentiable, thus compatible with gradient descent learning approaches via back-propagation.

\subsection{Limits of the p-convolution layer}\label{subsec:pconv_limit}

In order for the $PConv$ layer to be defined on all its possible input parameters, $w$ and $f$ must be strictly positive. Otherwise, the following issues can occur:
\begin{itemize}
    \item If $f(x)$ contains null values and $p$ is negative, $f^p(x)$ is not defined;
    \item If $f(x)$ contains negative values and $p$ is a non-null, non-integer real number, $f^p(x)$ can contain complex numbers;
    \item If $w(x)$ or $f^p(x)$ contain null values, $\frac{1}{(f^p * w)(x)}$ is not defined.
\end{itemize}
As such, before feeding an image to the p-convolution operation, it first must be rescaled between $[1, 2]$:
\begin{equation}
    \label{eq:rescale}
    f_r(x) = 1.0 + \frac{f(x) - \min_{x \in E} f(x)}{\max_{x \in E} f(x) - \min_{x \in E} f(x)}
\end{equation}
Moreover, if several $PConv$ layers are concatenated one behind the other (to achieve (pseudo-) opening and closing operations for instance), a rescaling must be performed before each layer.
Particular care must also be taken with the output of the last $PConv$ layer, since it must also be rescaled to ensure that the range of the output matches that of the target. This is done by adding a trainable scale/bias 1x1x1 convolution layer at the end of the network.\\
Last but not least, a notable drawback of the $PConv$ layer when it comes to learning a specific (binary or non-flat) structuring element is that it tends to be hollow and flattened out in the center (see further presented results in Section~\ref{sec:xp}).

\section{Proposed $\lmorph$ and $\smorph$ layers}\label{sec:contrib}

As exposed in the previous section~\ref{subsec:pconv_limit}, the $PConv$ layer has a few edge cases and drawbacks. For this reason, we now propose two new morphological layers, based upon the same fundamental principle as the $PConv$ operation, with the intent of making them compatible with general grayscale mathematical morphology.

\subsection{Introducing the $\lmorph$ operation}

\begin{figure}[tb]
\centering
\begin{minipage}{0.24\textwidth}
\centering
\includegraphics[width=0.98\textwidth]{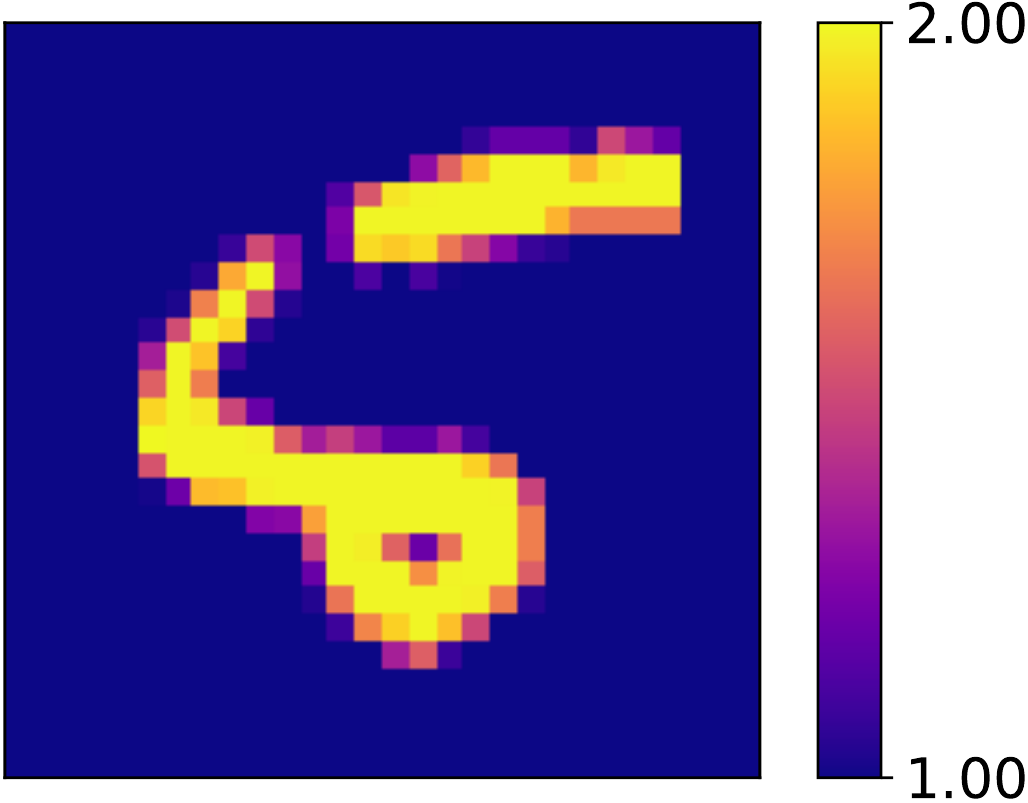}
\end{minipage}\hfill
\begin{minipage}{0.24\textwidth}
\centering
\includegraphics[width=0.98\textwidth]{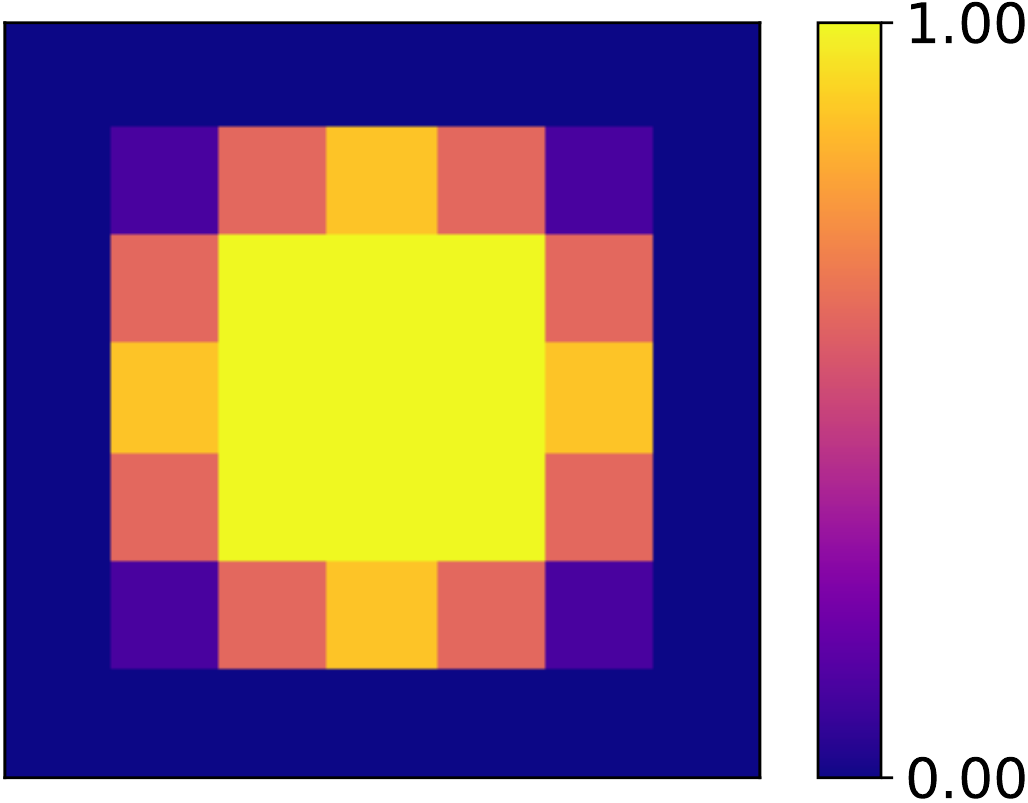}
\end{minipage}\hfill
\begin{minipage}{0.24\textwidth}
\centering
\includegraphics[width=0.98\textwidth]{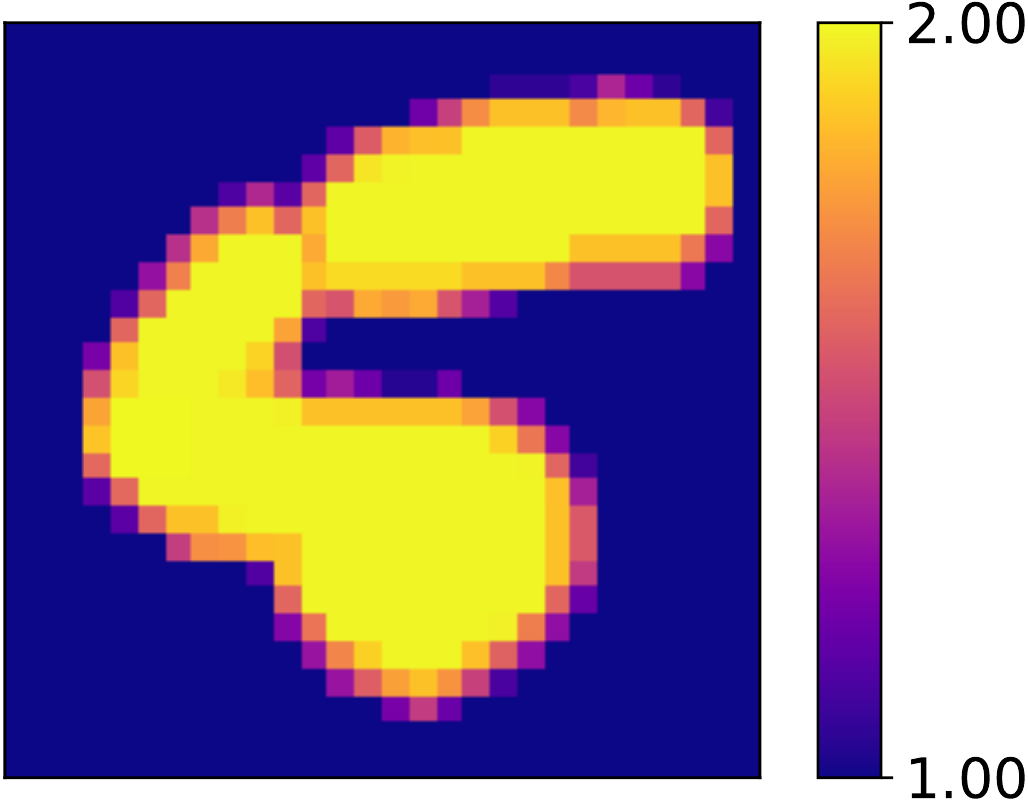}
\end{minipage}\hfill
\begin{minipage}{0.24\textwidth}
\centering
\includegraphics[width=0.98\textwidth]{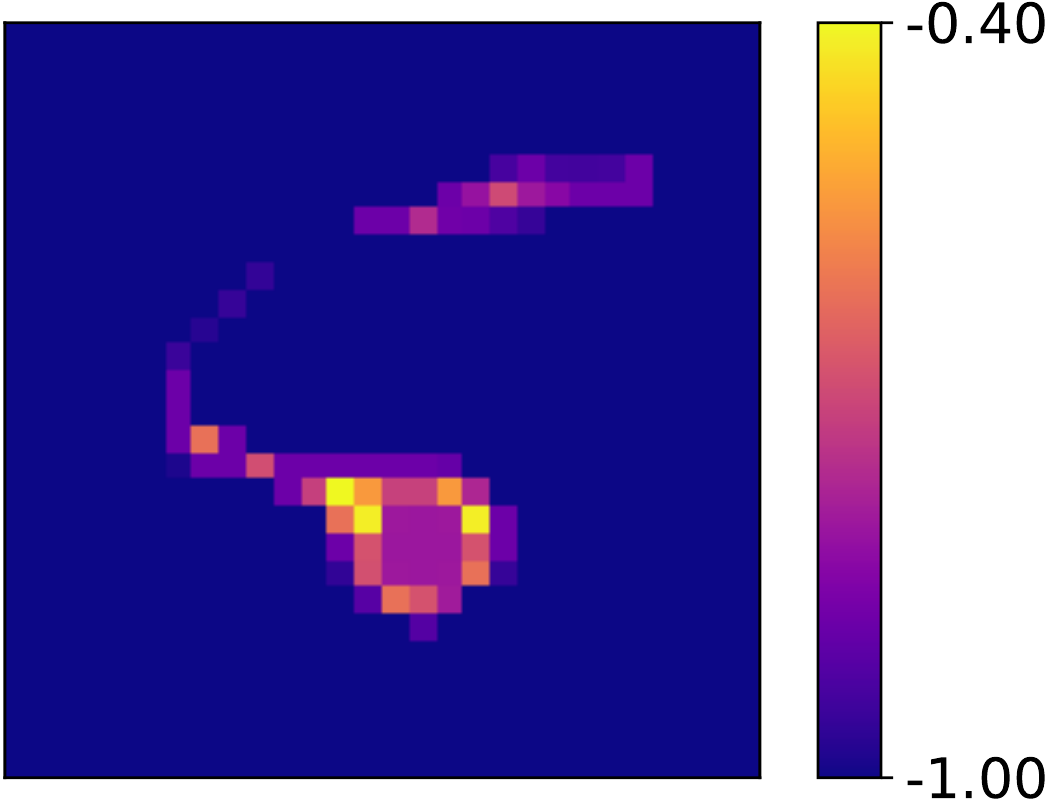}
\end{minipage}\\[5pt]
\begin{minipage}{0.24\textwidth}
\centering
\includegraphics[width=0.98\textwidth]{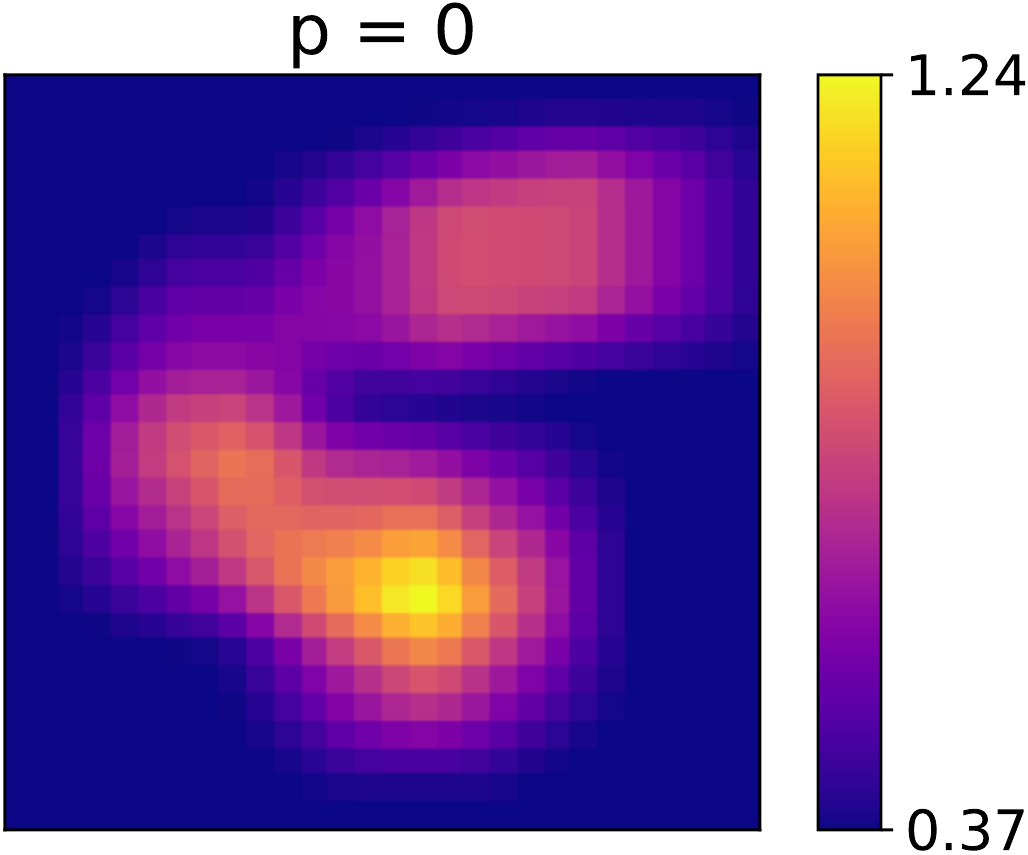}
\end{minipage}\hfill
\begin{minipage}{0.24\textwidth}
\centering
\includegraphics[width=0.98\textwidth]{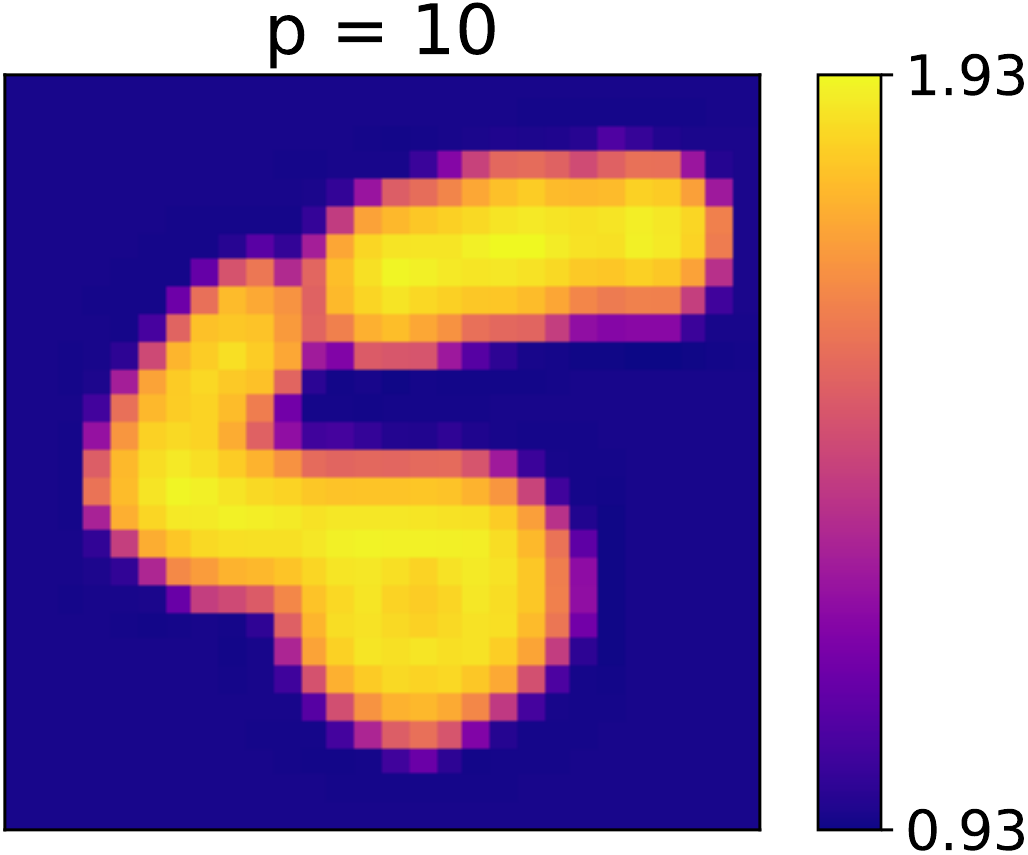}
\end{minipage}\hfill
\begin{minipage}{0.24\textwidth}
\centering
\includegraphics[width=0.98\textwidth]{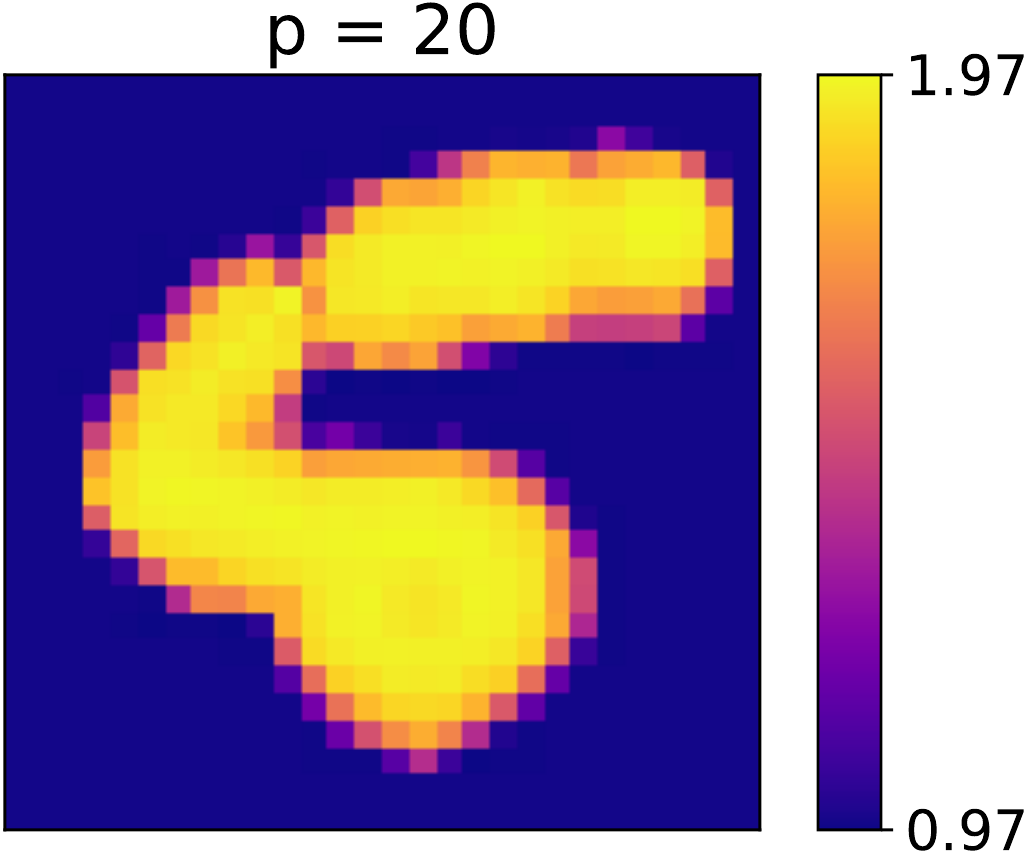}
\end{minipage}\hfill
\begin{minipage}{0.24\textwidth}
\centering
\includegraphics[width=0.98\textwidth]{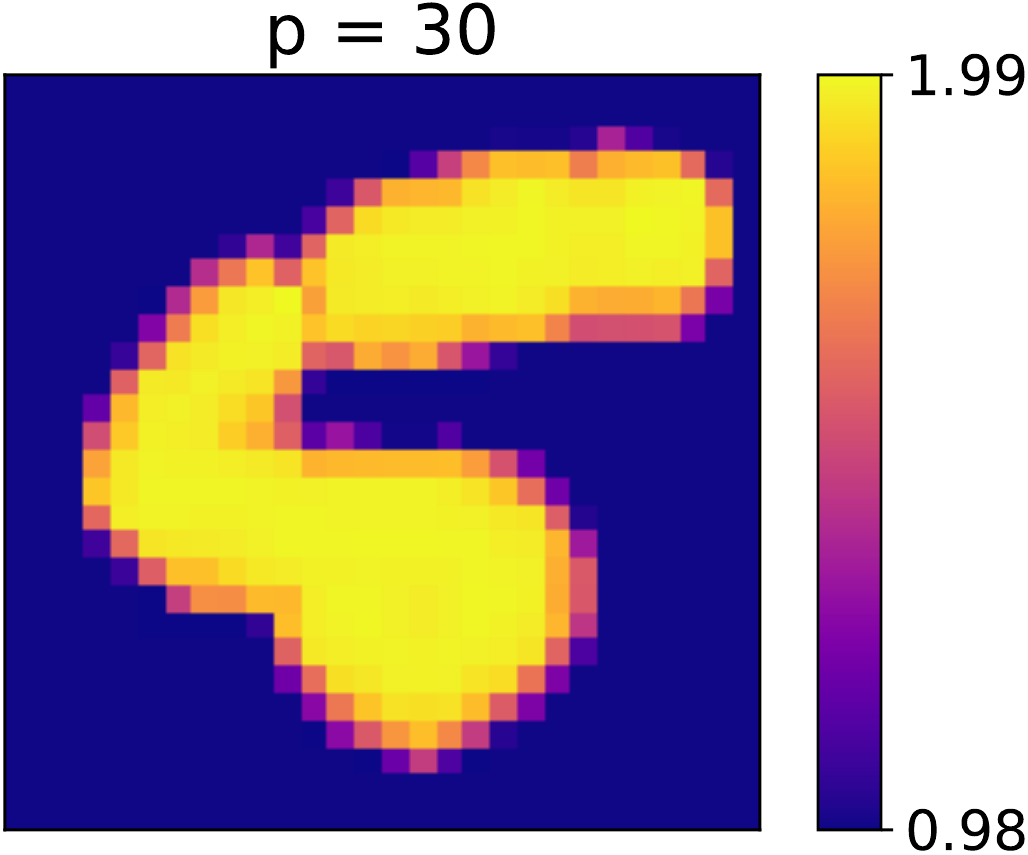}
\end{minipage}\\[5pt]
\begin{minipage}{0.24\textwidth}
\centering
\includegraphics[width=0.98\textwidth]{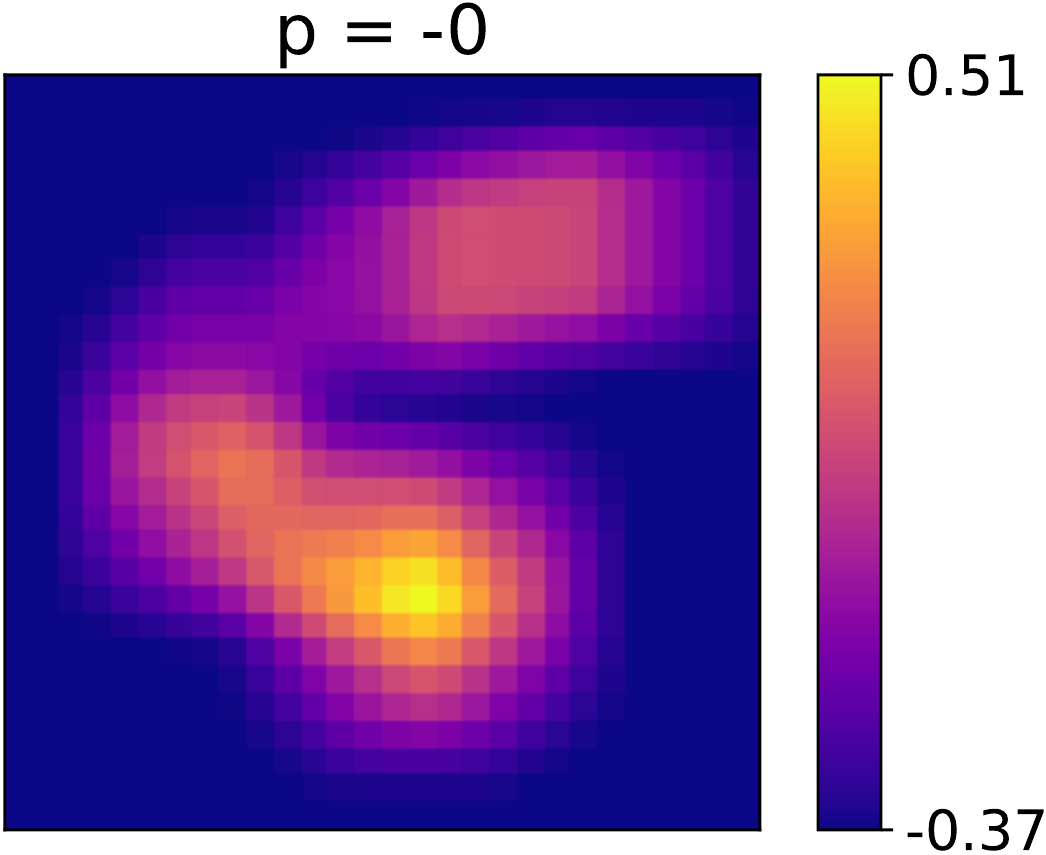}
\end{minipage}\hfill
\begin{minipage}{0.24\textwidth}
\centering
\includegraphics[width=0.98\textwidth]{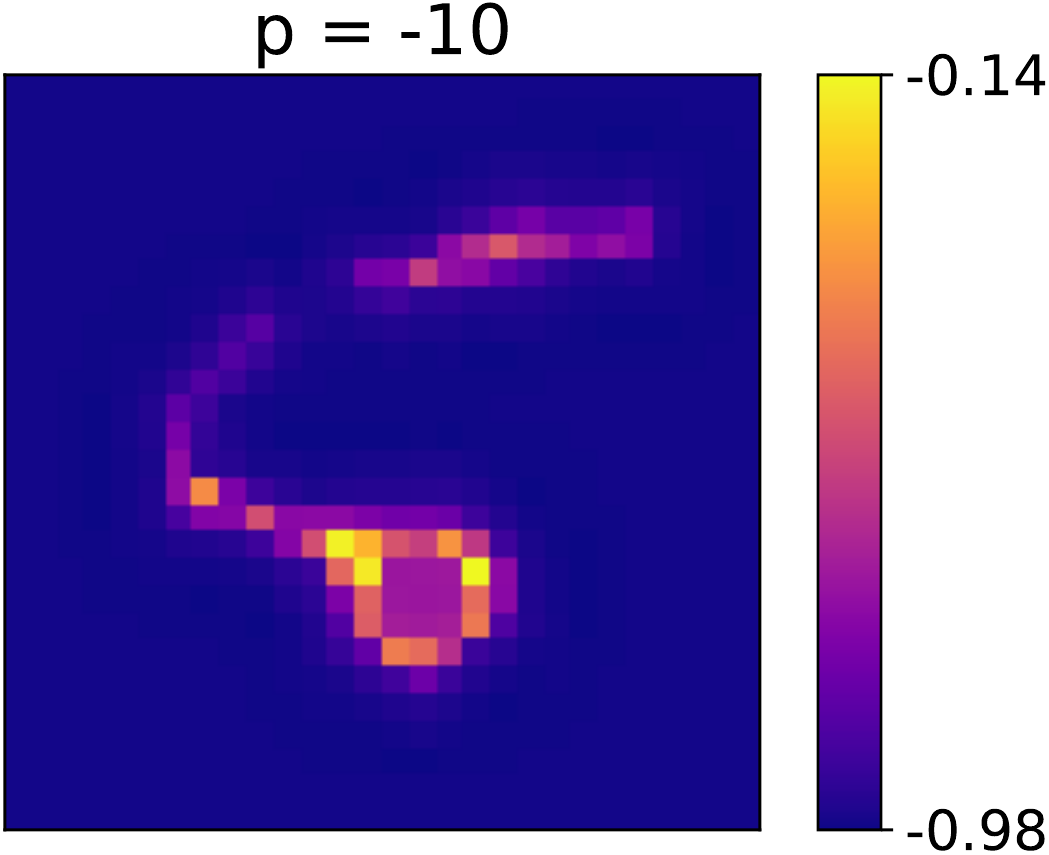}
\end{minipage}\hfill
\begin{minipage}{0.24\textwidth}
\centering
\includegraphics[width=0.98\textwidth]{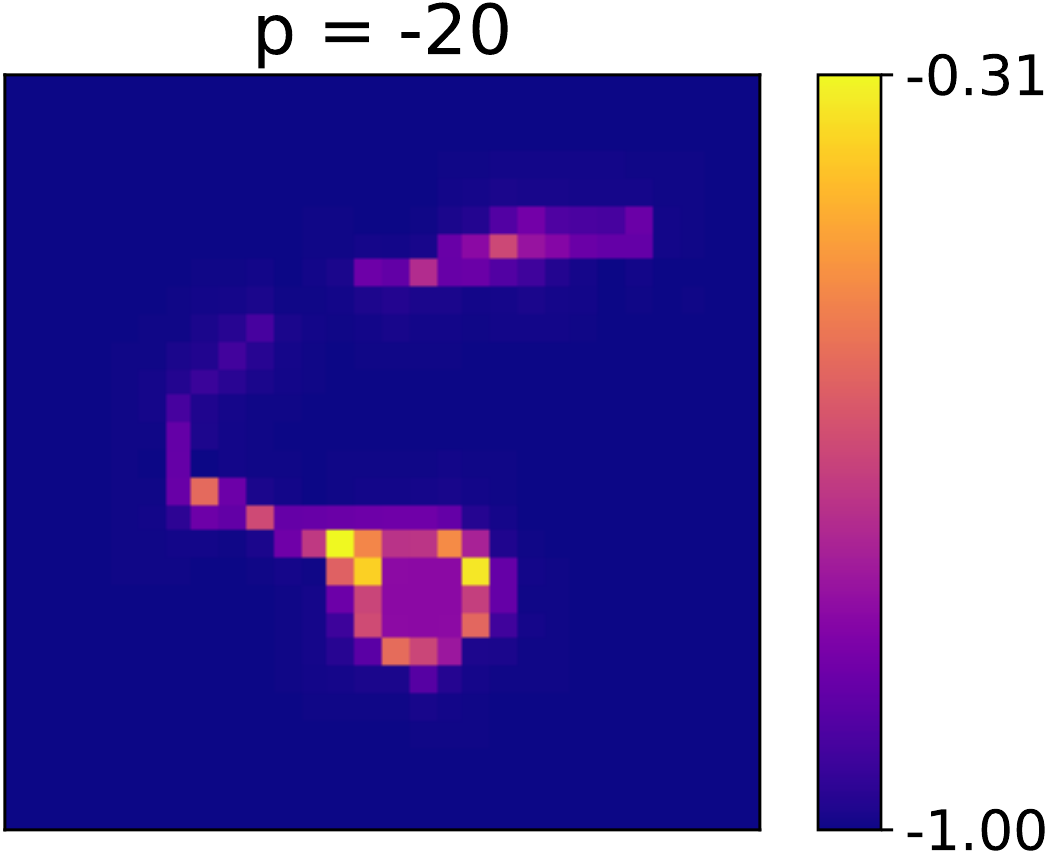}
\end{minipage}\hfill
\begin{minipage}{0.24\textwidth}
\centering
\includegraphics[width=0.98\textwidth]{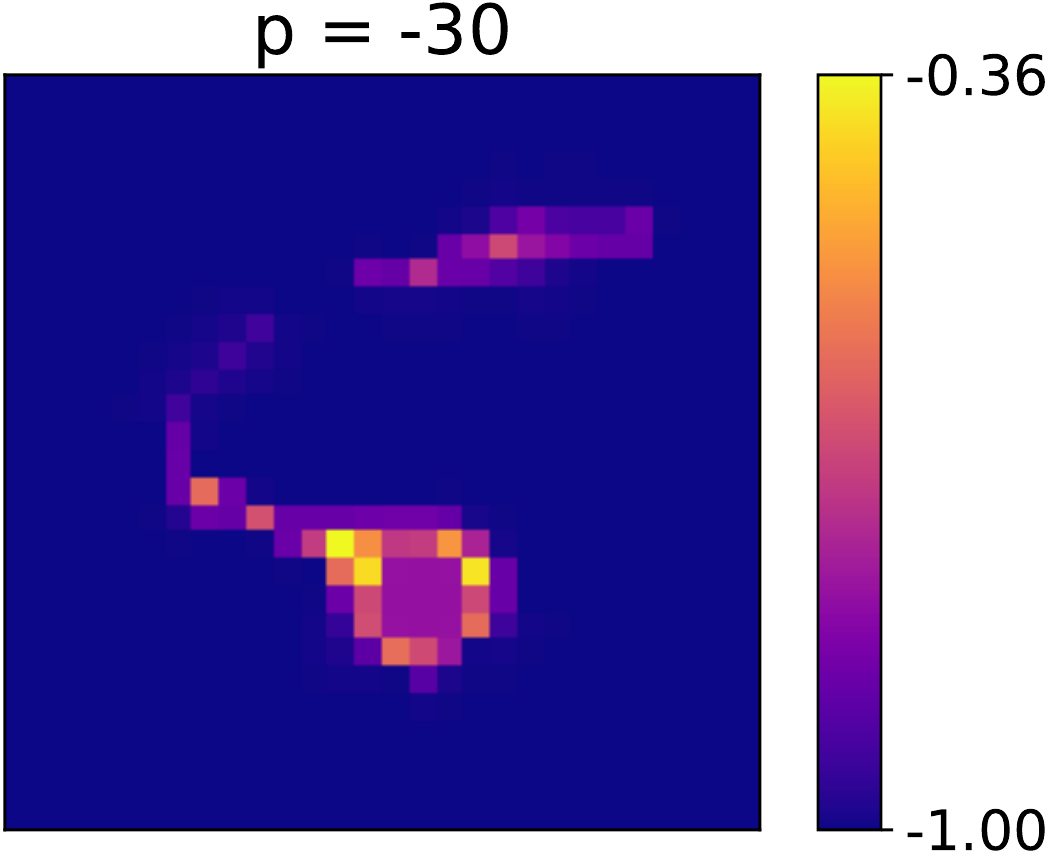}
\end{minipage}
\caption{Top row: input image, non-flat structuring element, target dilation, target erosion. Middle row: $\lmorph$ pseudo-dilation for increasing value of $p$. Bottom row: $\lmorph$ pseudo-erosion for increasing value of $\vert p \vert$.\vspace{-5pt}} \label{fig:lmorph_morpho}
\end{figure}

Our main objective is still to circumvent the non-differentiability of $\min$ and $\max$ functions by replacing them with smooth and differentiable approximations. In the previous section~\ref{sec:p-conv}, we presented the CHM, whose asymptotic behavior is exploited by the $PConv$ layer~\cite{masci2013learning}. Relying on this behavior once more, we now propose to define the following $\lmorph$ (for $\mathcal{L}$ehmer-mean based Morphological) operation:
\begin{equation}\label{eq:lmorph}
\lmorph(f, w, p)(x) = \frac{\sum _{{y \in W(x)}} \; (f(y) + w(x - y))^{p + 1}}{\sum _{{y \in W(x)}}  \;  (f(y) + w(x - y))^p}
\end{equation}
where $w : W \rightarrow \mathbb{R}^+$ is the structuring function and $p\in \mathbb{R}$. Defined as such, we can identify $\lmorph(f,w,p)$ with the CHM defined by the equation~\eqref{eq:CHM}: all weights $w_i$ (resp. entries $x_i$) of equation~\eqref{eq:CHM} correspond to $1$ (resp. $f(y) + w(x - y)$) in the equation~\eqref{eq:lmorph}, from which we can deduce the following asymptotic behavior:
\begin{align}
\lim_{p \rightarrow +\infty} \lmorph(f, w, p)(x) &= \sup_{y \in W(x)} \left\{f(y) + w(x-y) \right\}= (f \oplus w)(x) \label{eq:lmorph_dilation}\\
\lim_{p \rightarrow -\infty} \lmorph(f, w, p)(x) &= \inf_{y \in W(x)} \left\{f(y) + w(x-y) \right\}=(f \ominus -w)(x) \label{eq:lmorph_erosion}
\end{align}
By changing the sign of $p$, one can achieve either pseudo-dilation (if $p > 0$) or pseudo-erosion (if $p < 0$). Figure \ref{fig:lmorph_morpho} displays examples of applying the $\lmorph$ function with a given non-flat structuring element for different values of $p$. In practice $\vert p \vert > 20$ is sufficient to reproduce non-flat grayscale dilation or non-flat grayscale erosion. Note however that the applied structuring function is $-w$ in the case of an erosion.\\
Relying on the CHM like the $PConv$ layer brings over some shared limitations: the input image $f$ must be positive and rescaled following equation~\eqref{eq:rescale}, and the structuring function $w$ must be positive or null.

\subsection{Introducing the $\smorph$ operation}\label{subsec:smorph}

\begin{figure}[tb]
\centering
\begin{minipage}{0.24\textwidth}
\centering
\includegraphics[width=0.98\textwidth]{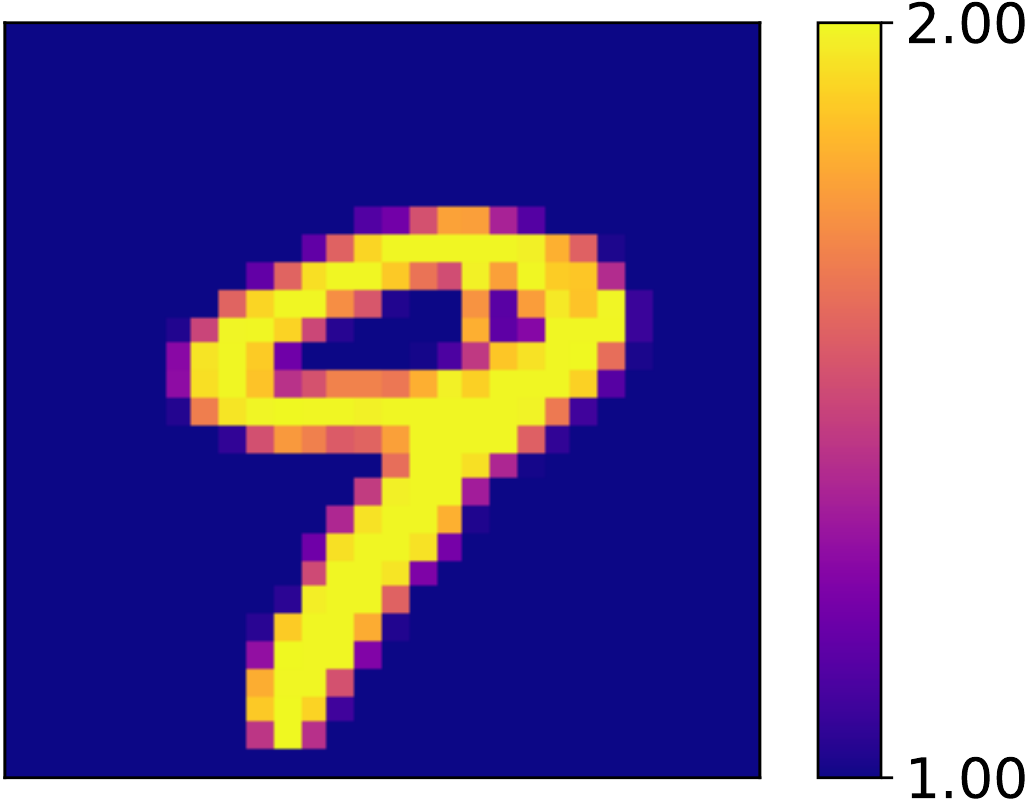}
\end{minipage}\hfill
\begin{minipage}{0.24\textwidth}
\centering
\includegraphics[width=0.98\textwidth]{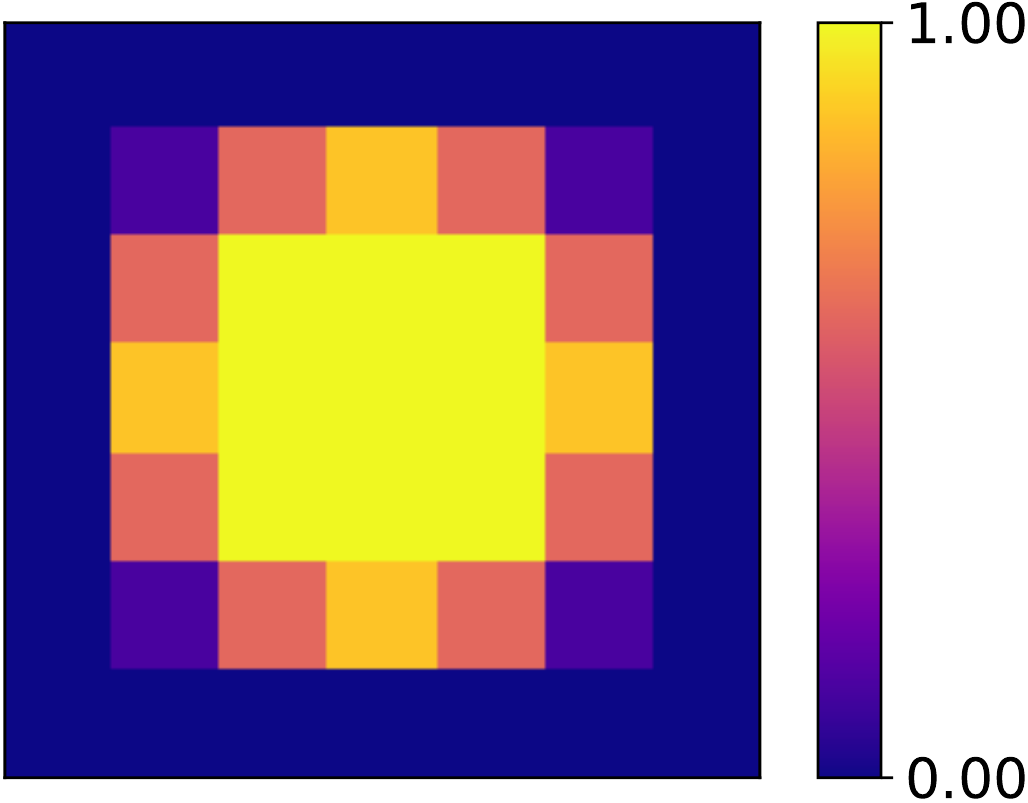}
\end{minipage}\hfill
\begin{minipage}{0.24\textwidth}
\centering
\includegraphics[width=0.98\textwidth]{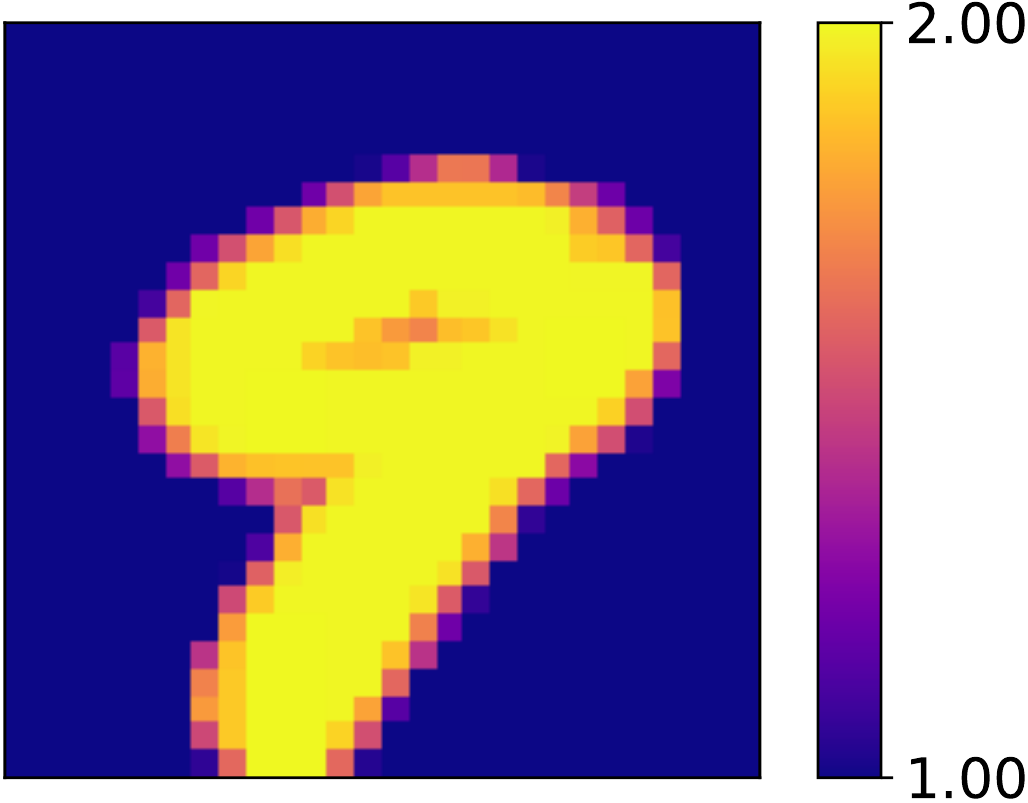}
\end{minipage}\hfill
\begin{minipage}{0.24\textwidth}
\centering
\includegraphics[width=0.98\textwidth]{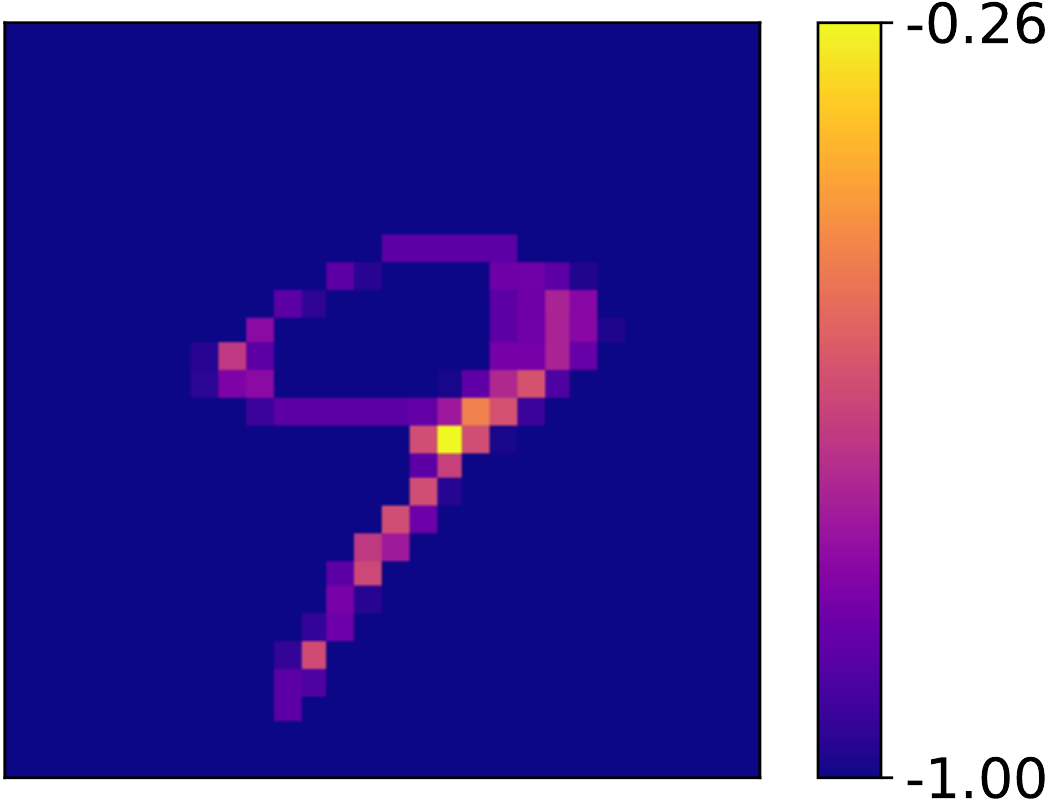}
\end{minipage}\\[5pt]
\begin{minipage}{0.24\textwidth}
\centering
\includegraphics[width=0.98\textwidth]{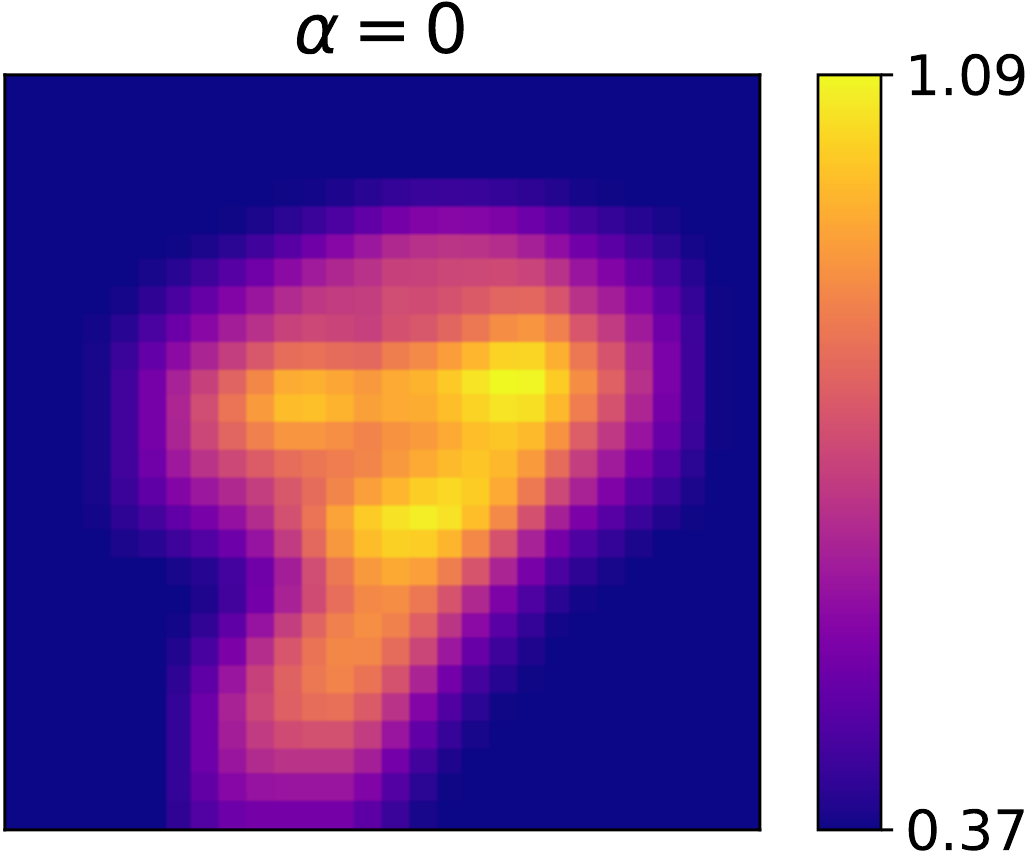}
\end{minipage}\hfill
\begin{minipage}{0.24\textwidth}
\centering
\includegraphics[width=0.98\textwidth]{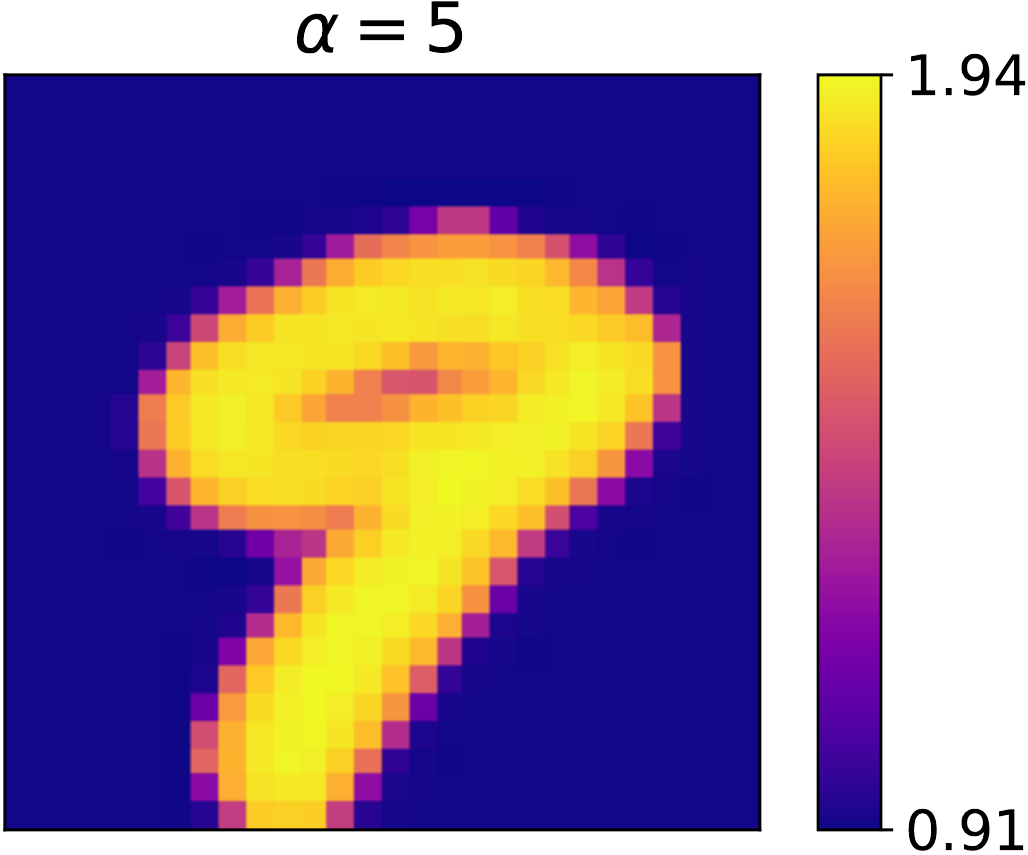}
\end{minipage}\hfill
\begin{minipage}{0.24\textwidth}
\centering
\includegraphics[width=0.98\textwidth]{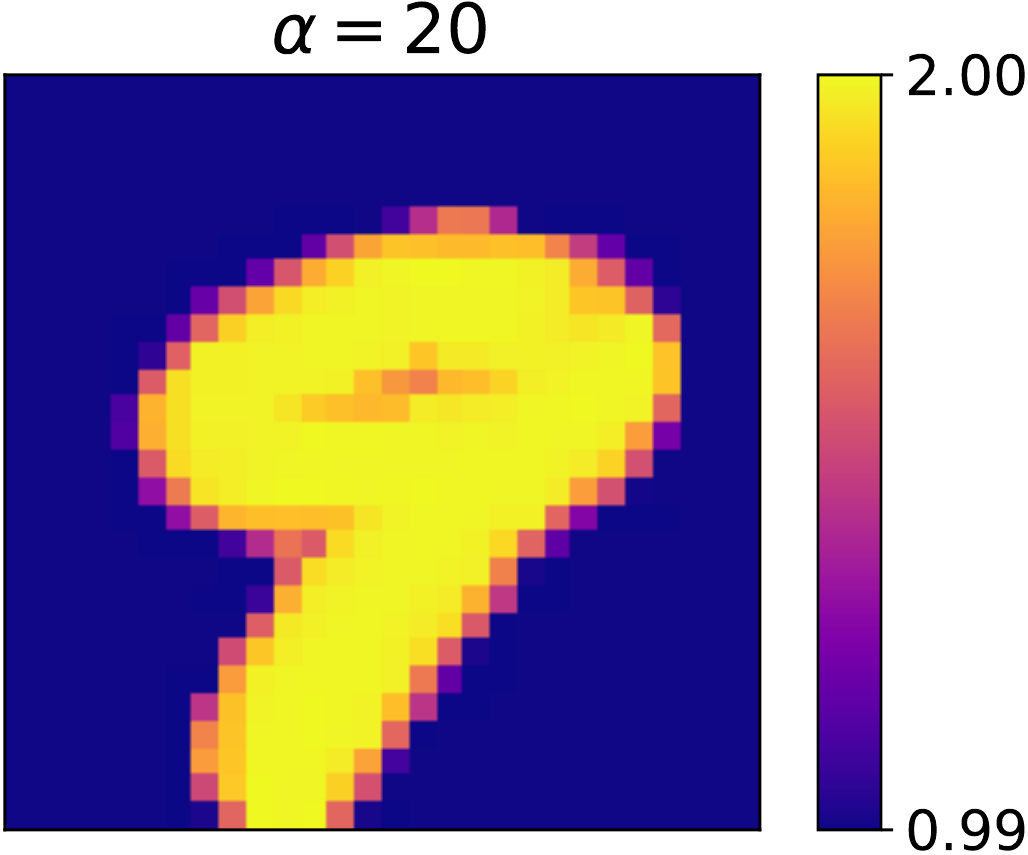}
\end{minipage}\hfill
\begin{minipage}{0.24\textwidth}
\centering
\includegraphics[width=0.98\textwidth]{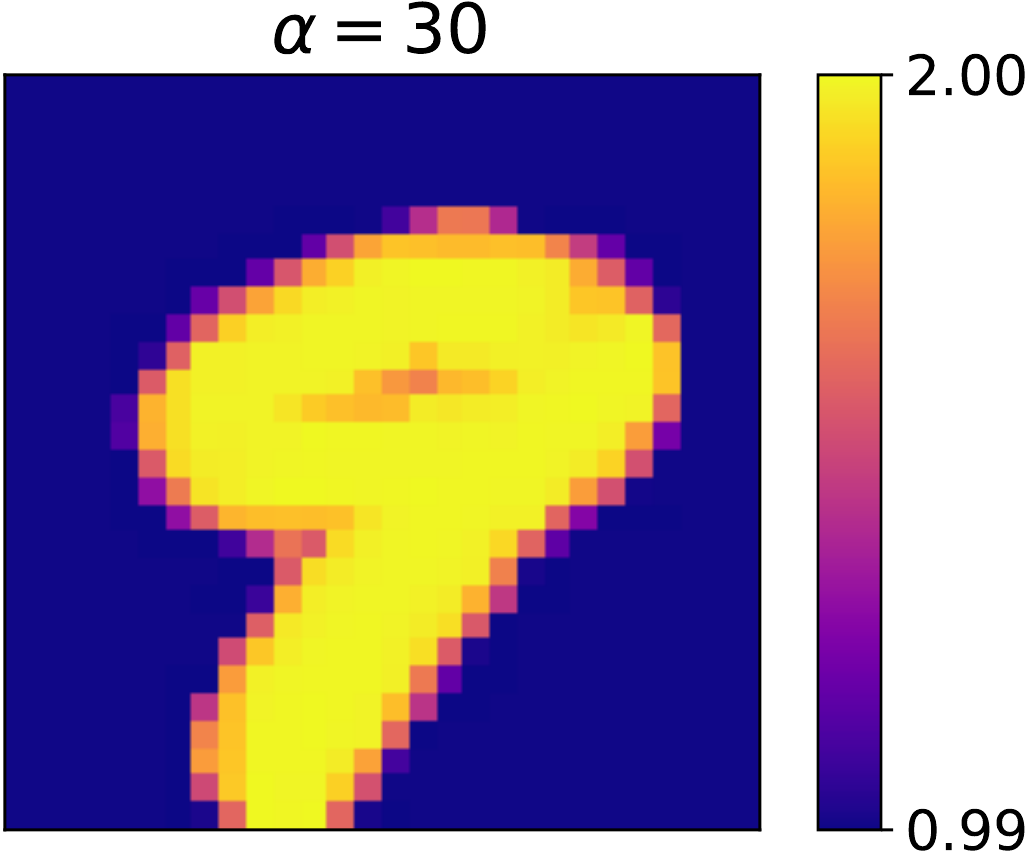}
\end{minipage}\\[5pt]
\begin{minipage}{0.24\textwidth}
\centering
\includegraphics[width=0.98\textwidth]{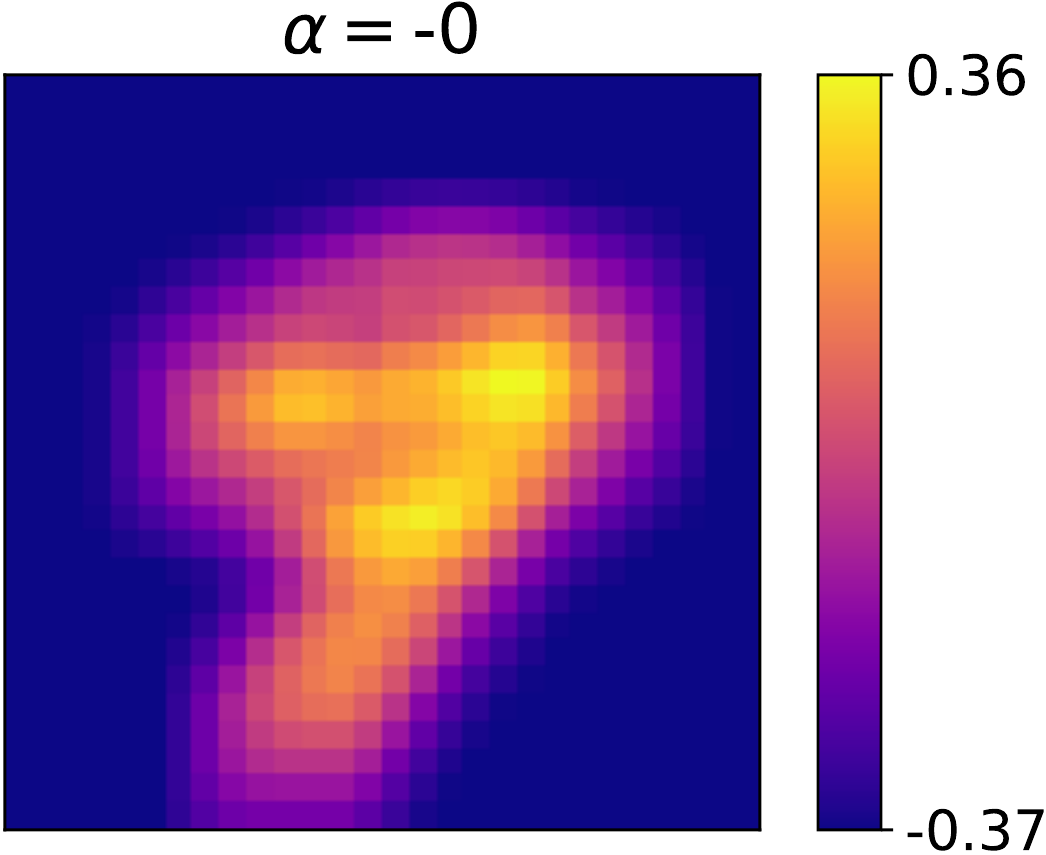}
\end{minipage}\hfill
\begin{minipage}{0.24\textwidth}
\centering
\includegraphics[width=0.98\textwidth]{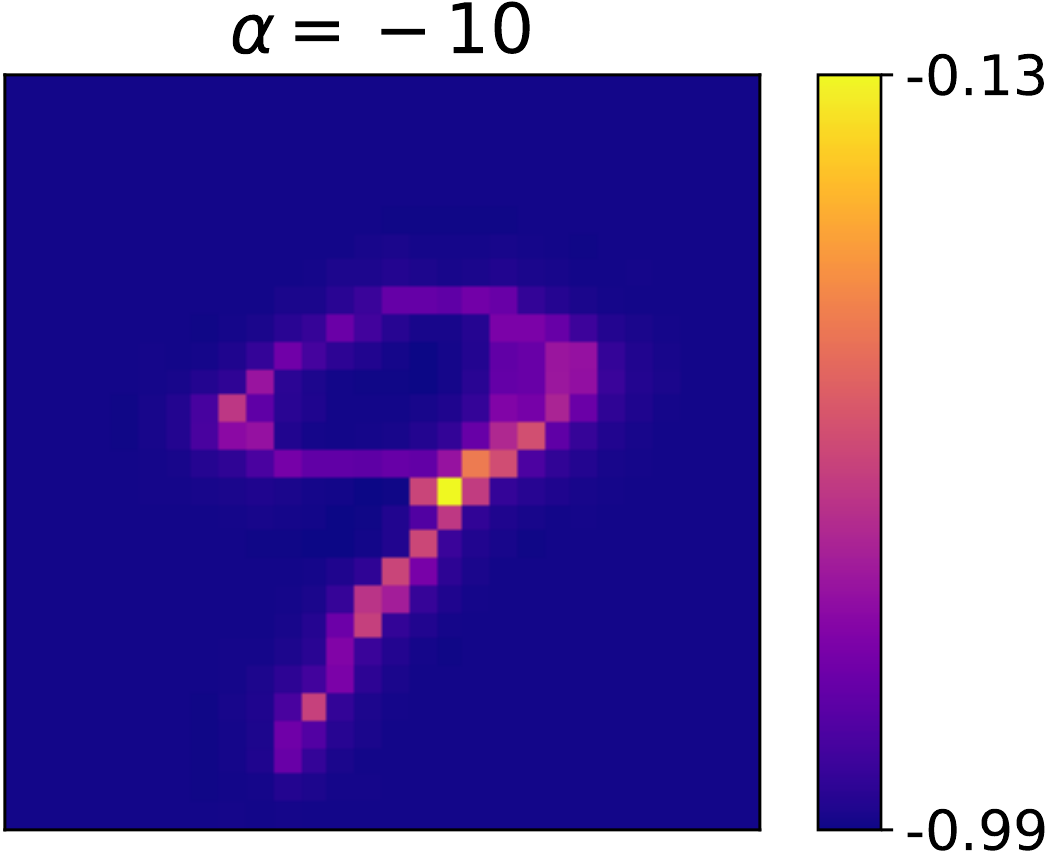}
\end{minipage}\hfill
\begin{minipage}{0.24\textwidth}
\centering
\includegraphics[width=0.98\textwidth]{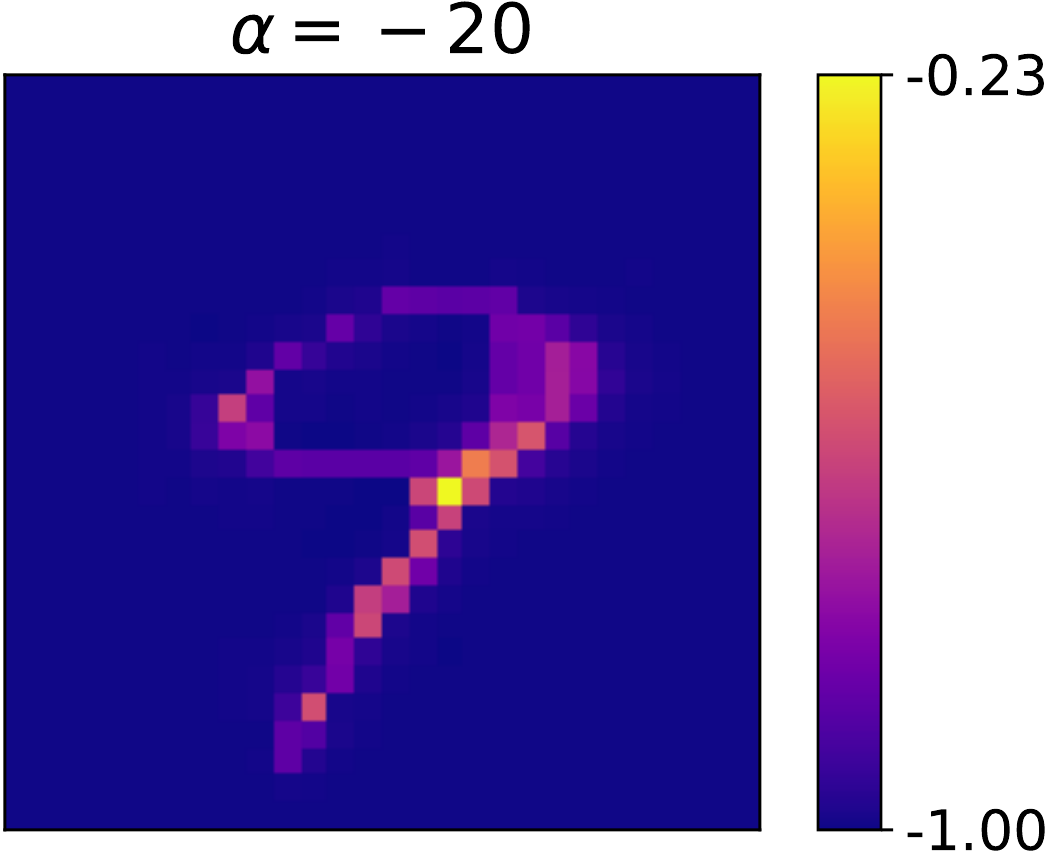}
\end{minipage}\hfill
\begin{minipage}{0.24\textwidth}
\centering
\includegraphics[width=0.98\textwidth]{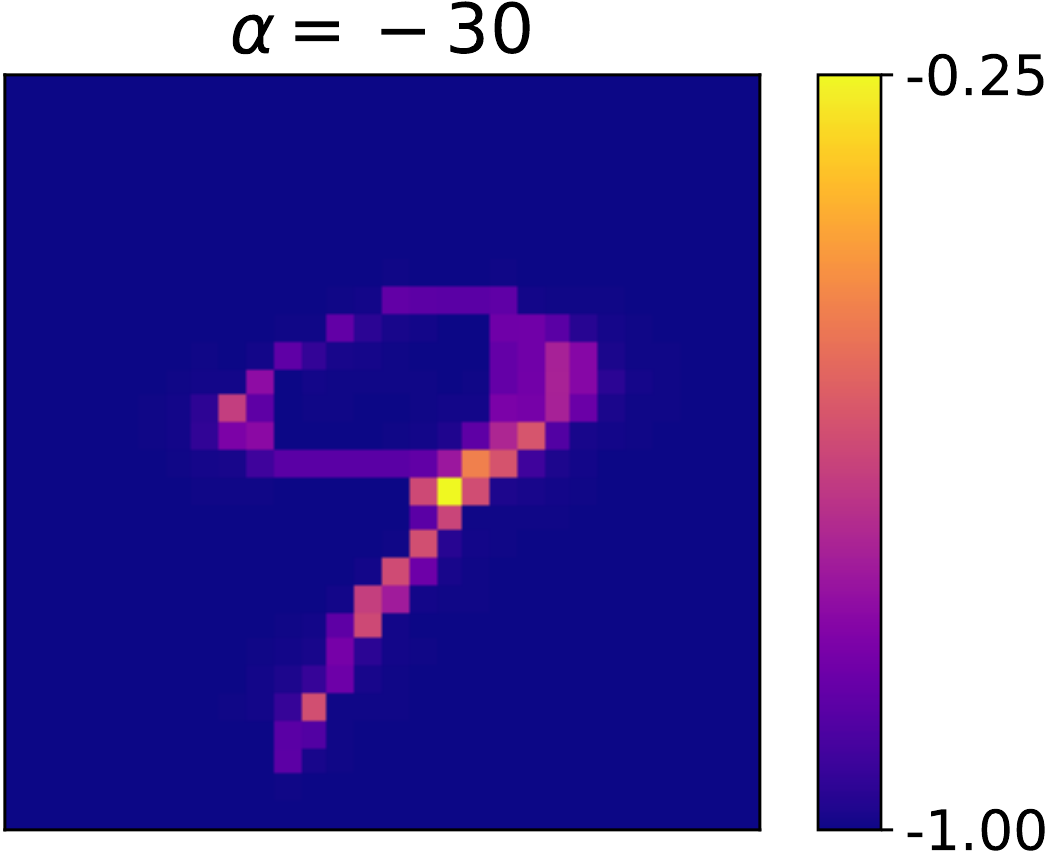}
\end{minipage}
\caption{Top row: input image, non-flat structuring element, target dilation, target erosion. Middle row: $\smorph$ pseudo-dilation for increasing value of $\alpha$. Bottom row: $\smorph$ pseudo-erosion for increasing value of $\vert \alpha \vert$.} \label{fig:smorph_morpho}
\end{figure}

Deriving a morphological layer based on the asymptotic behavior of the CHM has a major drawback in that the input must be rescaled within the range $[1, 2]$. In order to circumvent this issue, we now propose to leverage the $\alpha$-softmax function~\cite{lange2014applications}, which is defined as:
\begin{equation}\label{eq:alphasoftmax}
\mathcal{S}_\alpha(\mathbf{x}) = \frac{\sum^n_{i=1} x_i e^{\alpha x_i}}{\sum^n_{i=1} e^{\alpha x_i}}
\end{equation}
for some $\mathbf{x} = (x_1,\dots,x_n) \in \mathbb{R}^n$ and $\alpha \in \mathbb{R}$. In fact, $\mathcal{S}_\alpha$ has the desired properties that $\lim_{\alpha \rightarrow +\infty} \mathcal{S}_\alpha(\mathbf{x}) = \max_i x_i$ and $\lim_{\alpha \rightarrow -\infty} \mathcal{S}_\alpha(\mathbf{x}) = \min_i x_i$.
This function is less restrictive than the CHM since it does not require the elements of $\mathbf{x}$ to be strictly positive. A major benefit is that it is no longer necessary to rescale its input.\\
Exploiting this property, we define in the following the $\smorph$ (standing for Smooth Morphological) operation:
\begin{equation}
\smorph(f, w, \alpha)(x) = \frac{\sum _{{y \in W(x)}} (f(y) + w(x - y)) e^{\alpha (f(y) + w(x - y))}}{\sum _{{y \in W(x)}} e^{\alpha (f(y) + w(x - y))}},
\end{equation}
where $w : W \rightarrow \mathbb{R}$ plays the role of the structuring function.
We can see from the properties of $\mathcal{S}_\alpha$ that the following holds true:
\begin{align}
\lim_{\alpha \rightarrow +\infty} \smorph(f, w, \alpha)(x) &= (f \oplus w)(x) \label{eq:smorph_dilation}\\
\lim_{\alpha \rightarrow -\infty} \smorph(f, w, \alpha)(x) &= (f \ominus -w)(x)\label{eq:smorph_erosion}
\end{align}
As such, just like the $PConv$ and $\lmorph$ layers, the proposed $\smorph$ operation can alternate between a pseudo-dilation ($\alpha > 0$) and a pseudo-erosion ($\alpha < 0$). Furthermore, when $\alpha \gg 0$ (resp. $\alpha \ll 0$), this function approximates the grayscale dilation (resp. the grayscale erosion).\\
Figure \ref{fig:smorph_morpho} showcases examples of applying the $\smorph$ function with a given non-flat structuring element for different values of $\alpha$. We can see that, as $|\alpha|$ increases, the operation better and better approximates the target operation. 

\section{Conducted experiments}\label{sec:xp}

\subsection{Experimental protocol}\label{subsec:exp-protocol}

\begin{figure}[tb]
\centering
\begin{minipage}{0.13\textwidth}
\centering
\includegraphics[width=0.99\textwidth]{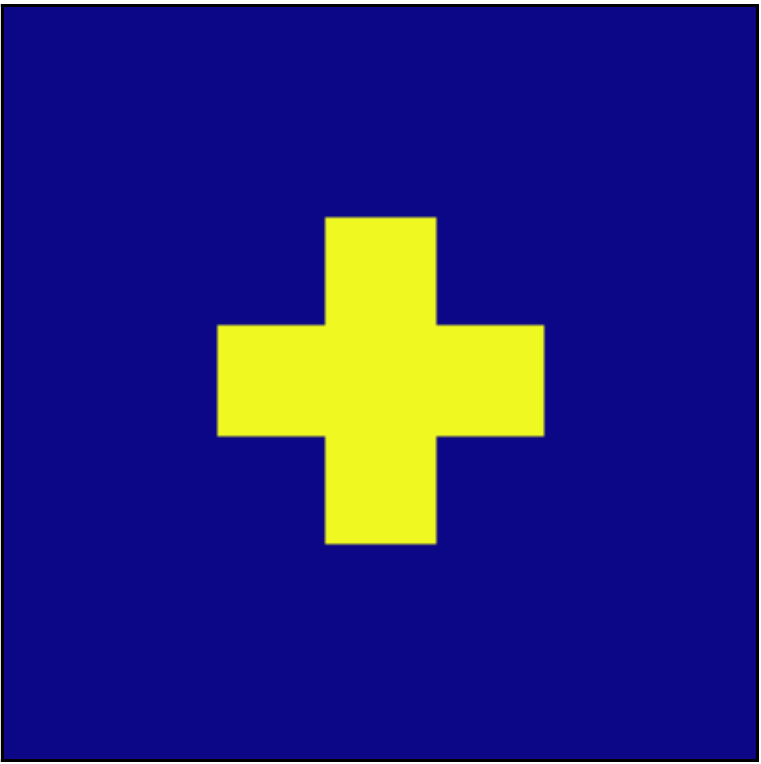}\\
\textit{cross3}
\end{minipage}\hfill
\begin{minipage}{0.13\textwidth}
\centering
\includegraphics[width=0.99\textwidth]{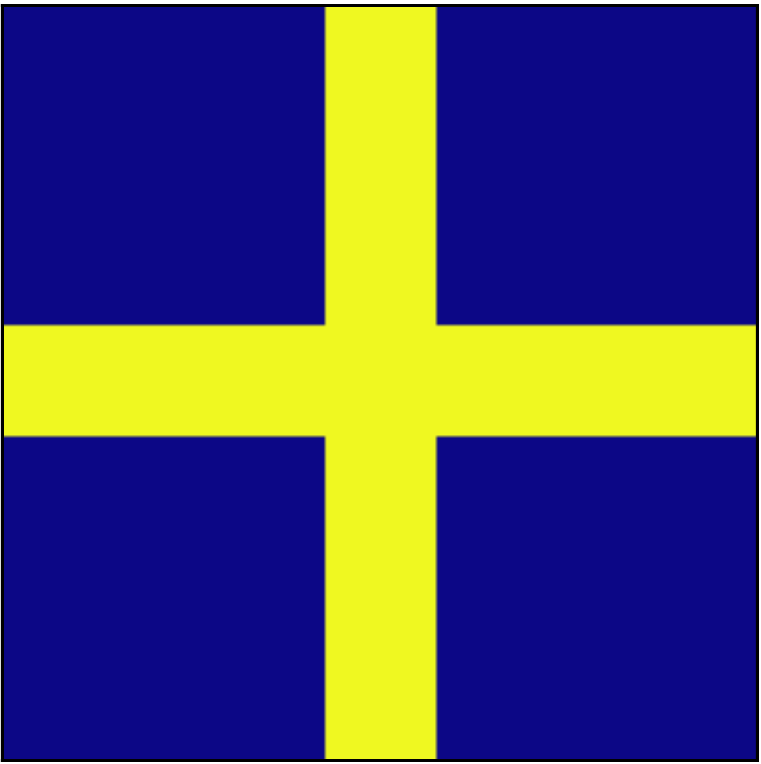}\\
\textit{cross7}
\end{minipage}\hfill
\begin{minipage}{0.13\textwidth}
\centering
\includegraphics[width=0.99\textwidth]{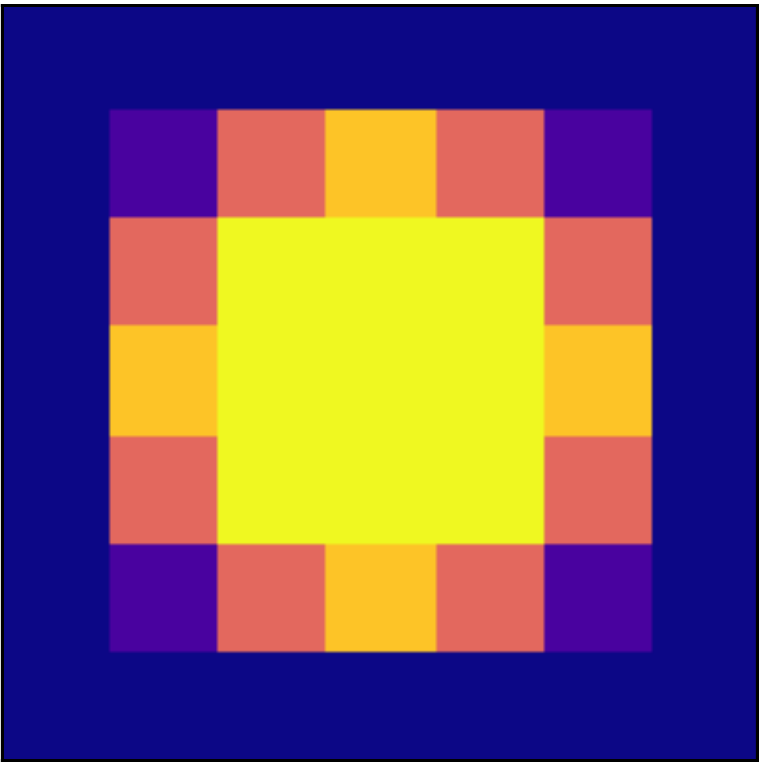}\\
\textit{disk2}
\end{minipage}\hfill
\begin{minipage}{0.13\textwidth}
\centering
\includegraphics[width=0.99\textwidth]{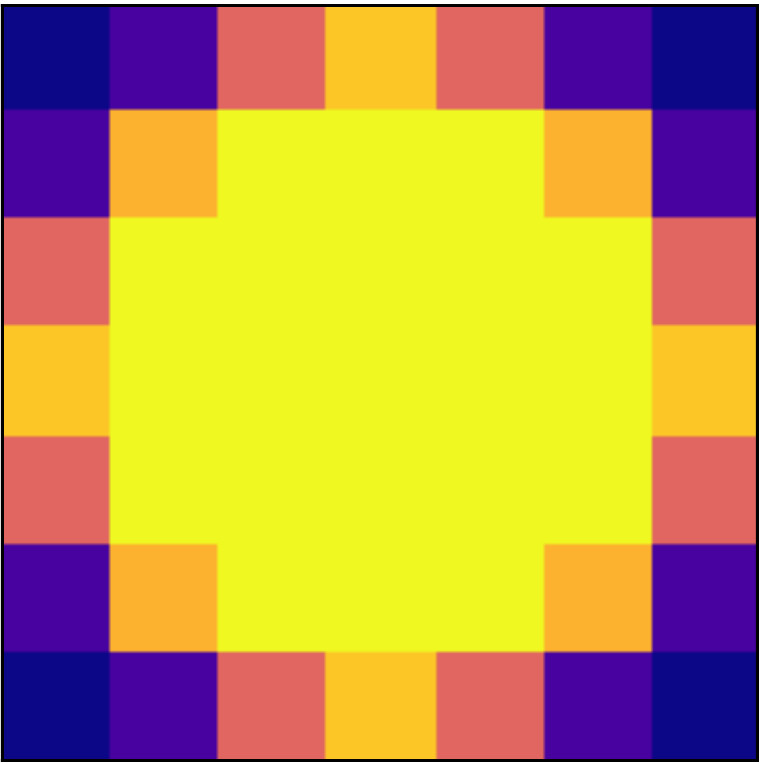}\\
\textit{disk3}
\end{minipage}\hfill
\begin{minipage}{0.13\textwidth}
\centering
\includegraphics[width=0.99\textwidth]{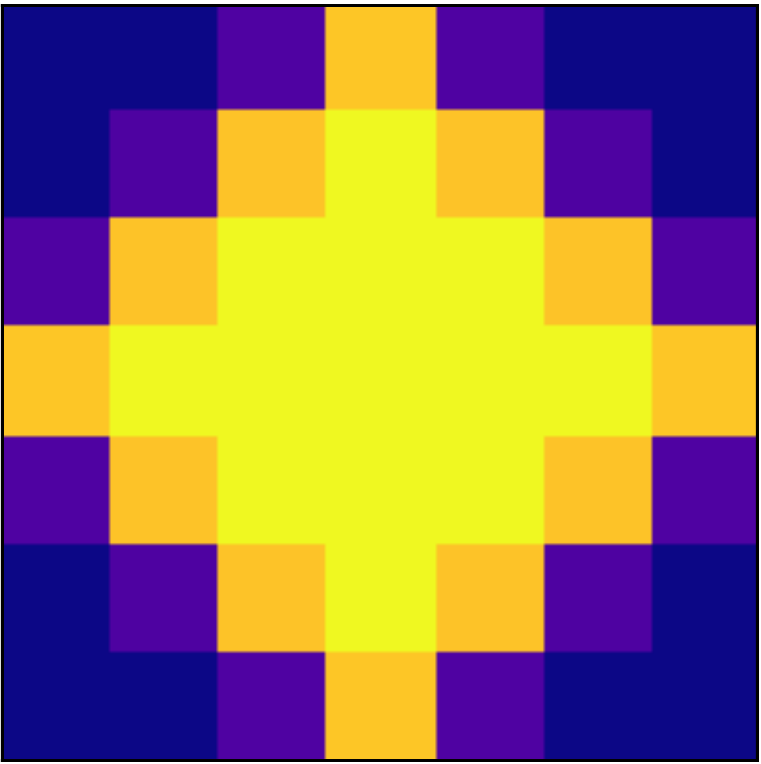}\\
\textit{diamond3}
\end{minipage}\hfill
\begin{minipage}{0.13\textwidth}
\centering
\includegraphics[width=0.99\textwidth]{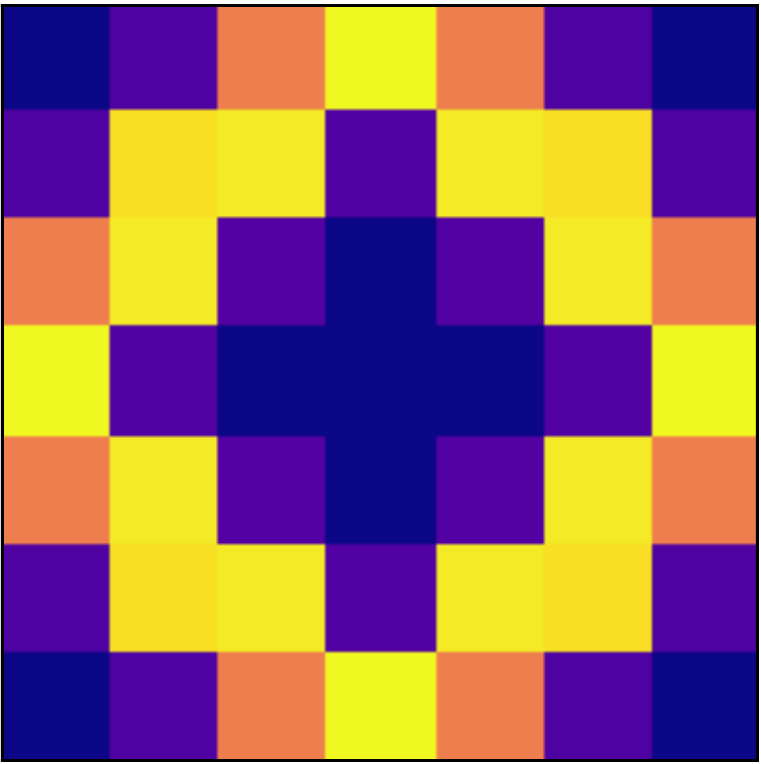}\\
\textit{complex}
\end{minipage}
\caption{$7 \times 7$ target grayscale structuring elements.}\label{fig:selems}
\vspace{-10pt}
\end{figure}

In the following, we evaluate the ability of the proposed $\lmorph$ and $\smorph$ layers to properly learn a target structuring element and compare with results obtained by the $PConv$ layer. To do so, we apply in turn dilation~$\oplus$, erosion~$\ominus$, closing~$\bullet$ and opening~$\circ$ to all 60000 digits of the MNIST dataset~\cite{lecun1998mnist}, with the target structuring elements displayed by Figure~\ref{fig:selems}. For dilation and erosion (resp. closing and opening), each network is composed of a single (resp. two) morphological layer(s) followed by a scale/bias $Conv$ 1x1x1 to rescale the output into the range of the target images. Note that for both $PConv$ and $\lmorph$ networks, the image also has to be rescaled in the range $[1,2]$ before passing through the morphological layer.
We train all networks with a batch size of 32, optimizing for the mean squared error (MSE) loss with the Adam optimizer (with starting learning rate $\eta = 0.01$). The learning rate of the optimizer is scheduled to decrease by a factor of $10$ when the loss plateaus for 5 consecutive epochs. Convergence is reached when the loss plateaus for 10 consecutive epochs.
For the $PConv$ layer, the filter is initialized with $1$s and $p=0$. For $\lmorph$, the filter is initialized with a folded normal distribution with standard deviation $\sigma = 0.01$, and $p=0$. For the $\smorph$ layer, the filter is initialized with a centered normal distribution with standard deviation $\sigma = 0.01$ and $\alpha = 0$. In all instances, the training is done simultaneously on the weights and the parameter $p$ or $\alpha$.\\
In order to assess the performance of the morphological networks for all scenarios (one scenario being one morphological operation $\oplus$, $\ominus$, $\bullet$ and $\circ$ and one target structuring element among those presented by Figure~\ref{fig:selems}), we computed the root mean square error (RMSE) between the filter learned at convergence and the target filter. The loss at convergence as well as the value of the parameter $p$ or $\alpha$ also serve as quantitative criteria.

\begin{figure}[tb]
\centering
\includegraphics[width=\textwidth]{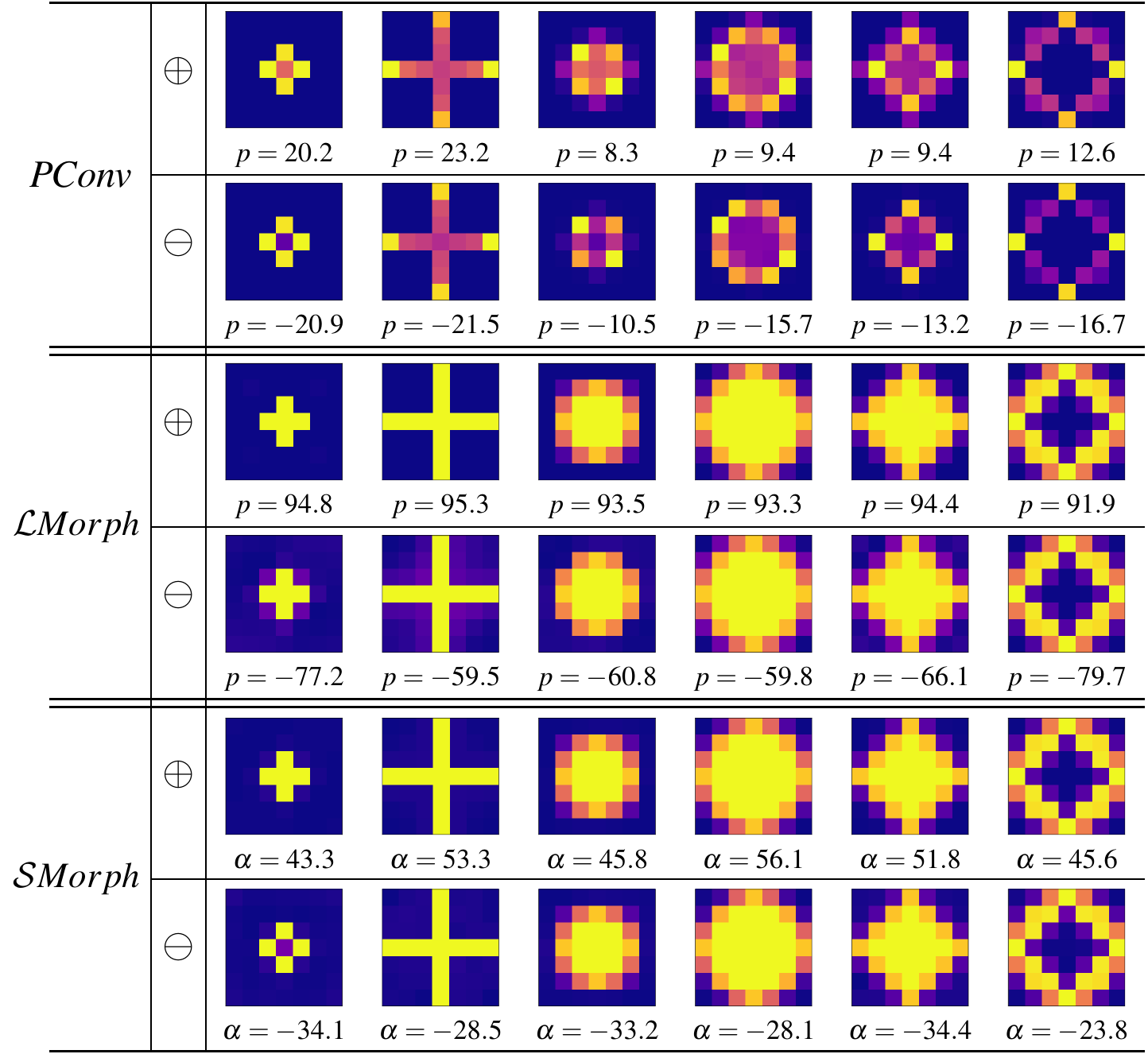}
\caption{Learned structuring element (with corresponding $p$ or $\alpha$ at convergence) for $PConv$, $\lmorph$ and $\smorph$ layers on dilation $\oplus$ and erosion $\ominus$ tasks.}\label{fig:results_dilation_erosion}
\end{figure}

\subsection{Obtained results}\label{subsec:results}

Figure~\ref{fig:results_dilation_erosion} gathers the structuring elements learned by the $PConv$, $\lmorph$ and $\smorph$ layers for dilation and erosion, and the value of their respective parameter. Looking at the sign of the parameter, all three networks succeed at finding the correct morphological operation. The magnitude of the parameter at convergence also confirms that the operation applied by all networks can be considered as dilation or erosion (and not simply pseudo-dilation or pseudo-erosion). However, looking at the shape of the learned structuring element, it is clear that the $PConv$ layer suffers from the hollow effect mentionned in section~\ref{subsec:pconv_limit}, while both $\lmorph$ and $\smorph$ layers accurately retrieve the target structuring element. This is confirmed by the RMSE values between those structuring elements and their respective targets, as presented by Table~\ref{tab:benchs_dilation_erosion}. More particularly, $\lmorph$ always achieves the lowest RMSE value for dilation tasks, while $\smorph$ succeeds better on erosion. In any case, the loss at convergence of the $\smorph$ network is almost consistently lower than that of the $\lmorph$ network by one or two orders of magnitude, and by two to three with respect to the $PConv$ network.
\begin{table}[tb]
\centering
\caption{MSE loss at convergence and RMSE between the learned structuring element displayed by Figure~\ref{fig:results_dilation_erosion} and the target for $PConv$, $\lmorph$ and $\smorph$ layers on dilation $\oplus$ and erosion $\ominus$ tasks. Best (lowest) results are in bold.}\label{tab:benchs_dilation_erosion}
\includegraphics[width=\textwidth]{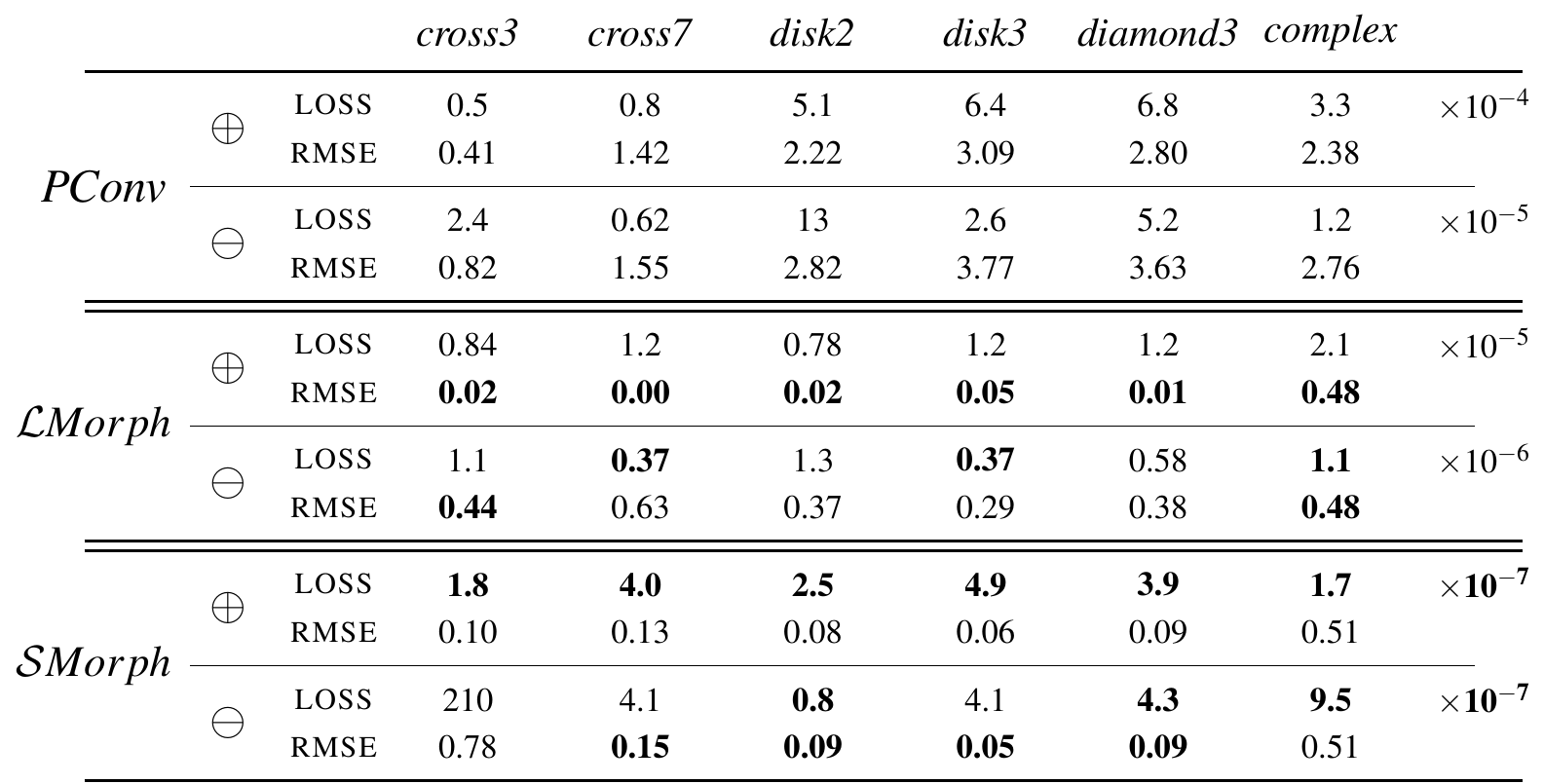}
\end{table}

\begin{figure}[tb]
\centering
\includegraphics[width=\textwidth]{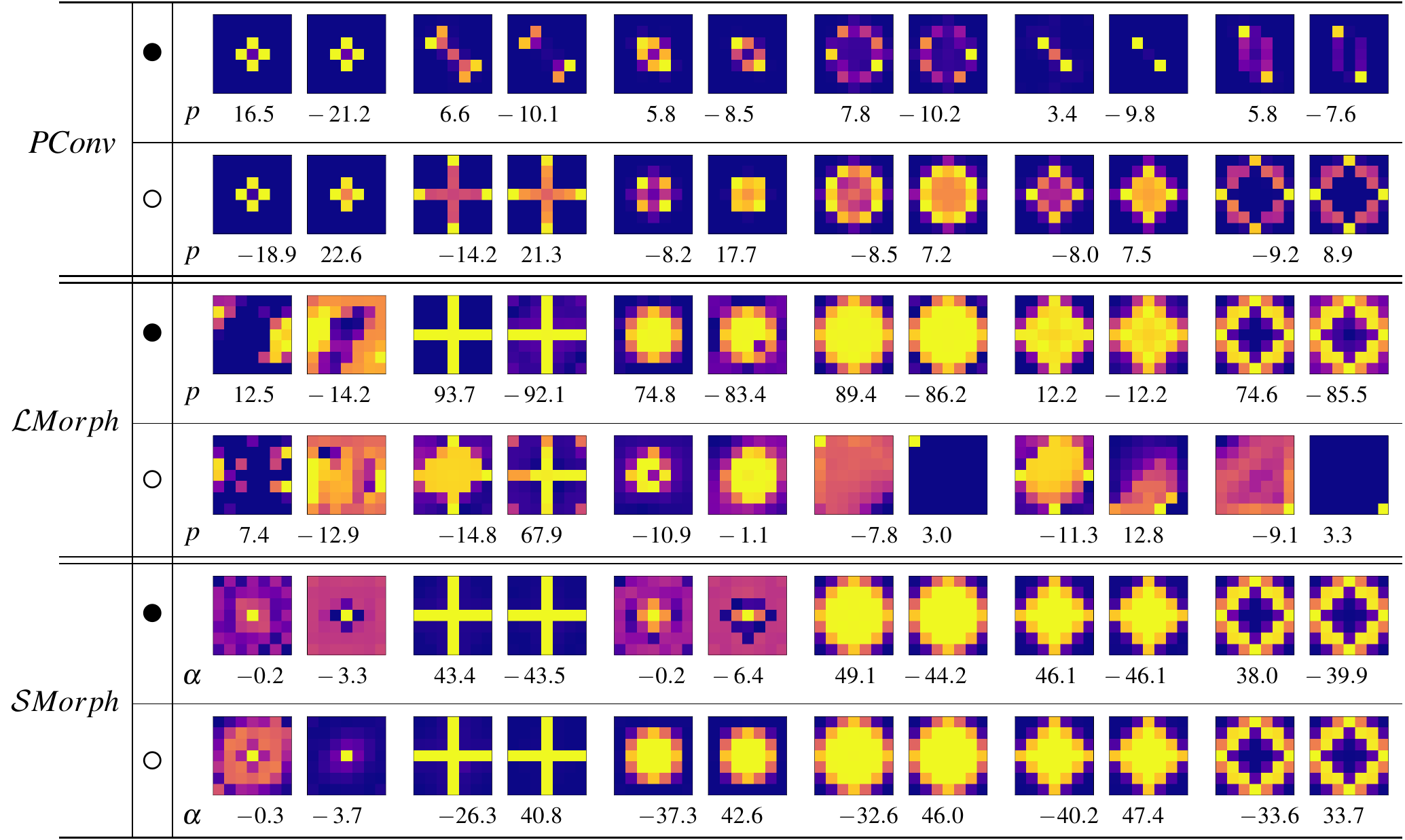}
\caption{Learned structuring elements (with corresponding $p$ or $\alpha$ value for each layer) for $PConv$, $\lmorph$ and $\smorph$ layers on closing $\bullet$ and opening $\circ$ tasks.}\label{fig:results_closing_opening}
\end{figure}
\noindent Figure~\ref{fig:results_closing_opening} displays the structuring elements learned by the $PConv$, $\lmorph$ and $\smorph$ layers for all 6 target structuring elements for closing and opening operations. This time, since each network comprises of two morphological layers (and a scale/bias $Conv$ 1x1x1 layer), it is worth mentioning that the two filters evolve independently from each other. Nevertheless, the two morphological layers are expected to learn filters having exactly the same shape, with parameter $p$ or $\alpha$ of opposite signs, once training converges.\\
As can be seen on Figure~\ref{fig:results_closing_opening}, the $PConv$ network always succeeds at learning the right morphological operation: the first (resp. second) layer converges to $ p > 0$ (resp. $p<0$) for the closing, and the opposite behavior for the opening. However, quite often $\vert p \vert < 10$, indicating that the layer is applying pseudo-dilation or pseudo-erosion only. In addition, the learned structuring element suffers again from the hollow effect for the opening, and does not find the correct shape for the closing.
The $\lmorph$ network succeeds in learning the correct operation and shape for the closing operation with large target structuring elements (all but \textit{cross3} and \textit{disk2}). For the opening operation however, it consistently fails at retrieving the shape of the target structuring element. This counter-performance is up to now unexplained.
The $\smorph$ network also struggles with small structuring elements for both the opening and closing, but perfectly recovers large ones.
The edge case of small target structuring elements could come from the scale/bias $Conv$ 1x1x1, which over-compensates for the gain or loss of average pixel intensities. Thus, when back-propagating the error during the learning phase, all filters behind the scale/bias $Conv$ 1x1x1 layer might not learn the right operation with a parameter $p$ or $\alpha$ not converging toward the correct sign domain.\\
Table~\ref{tab:benchs_closing_opening} presents the MSE loss at convergence and RMSE value between the learned filters and the target structuring elements for closing and opening scenarios. Except for the aforementioned edge case, the $\smorph$ layer consistently achieves the lowest loss value and RMSE for opening, while the best results for closing are obtained either by $\lmorph$ or $\smorph$ layers. Overall, apart from small structuring elements for closing or opening operations, the proposed $\smorph$ layer outperforms its $PConv$ and $\lmorph$ counterparts. Last but not least, it should also be noted that the $\smorph$ layer is also numerically more stable. As a matter of fact, raising to the power of $p$ in the $PConv$ and $\lmorph$ layers faster induces floating point accuracy issues.
\begin{table}[tb]
\centering
\caption{MSE loss at convergence and RMSE between the learned structuring elements displayed by Figure~\ref{fig:results_closing_opening} and the target for $PConv$, $\lmorph$ and $\smorph$ layers on closing $\bullet$ and opening $\circ$ tasks. Best (lowest) results are in bold.}\label{tab:benchs_closing_opening}
\includegraphics[width=\textwidth]{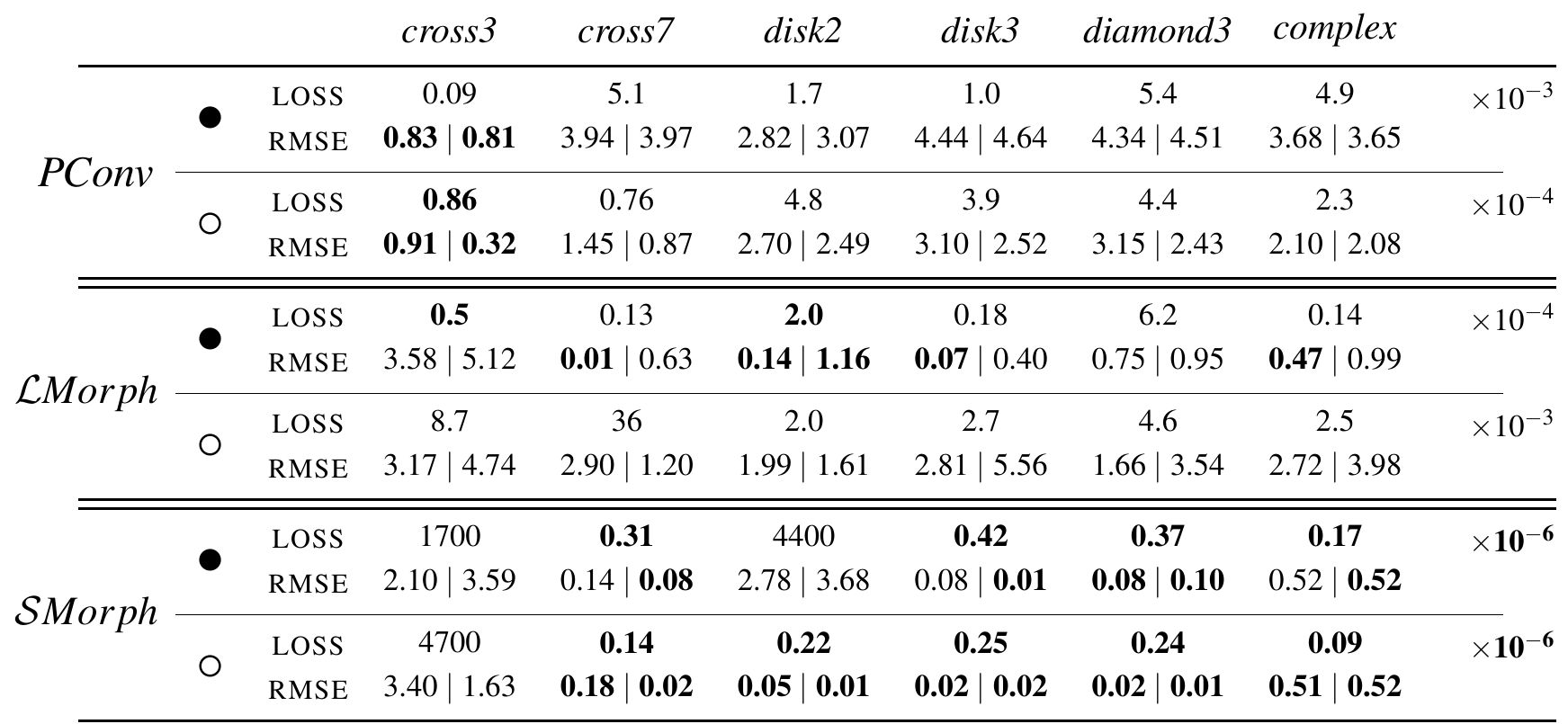}
\end{table}


\section{Conclusion}\label{sec:concl}

We present two new morphological layers, namely $\lmorph$ and $\smorph$. Similarly to the $Pconv$ layer of Masci \textit{et al.}~\cite{masci2013learning}, the former relies on the asymptotic properties of the CHM to achieve grayscale erosion and dilation. The latter instead relies on the $\alpha$-softmax function to reach the same goal, thus sidestepping some of the limitations shared by the $PConv$ and $\lmorph$ layers (namely being restricted to strictly positive inputs, rescaled in the range $[1,2]$, and positive structuring functions). In order to evaluate the performances of the proposed morphological layers, we applied in turn dilation, erosion, closing and opening on the whole MNIST dataset images, with $6$ target structuring elements of various sizes and shapes, and the morphological layers were trained to retrieve those structuring elements. Qualitative and quantitative comparisons demonstrated that the $\smorph$ layer overall outperforms both the $PConv$ and $\lmorph$ layers.\\
Future work includes investigating the edge cases uncovered for both proposed layers, as well as integrating them into more complex network architectures and evaluating them on concrete image processing applications.

\bibliographystyle{splncs04}
\bibliography{kirszie.2021.dgmm}

\end{document}